\documentstyle[12pt,a4]{article}

\begin{document}

\reversemarginpar

\title{Will small particles exhibit Brownian motion\\
in the quantum vacuum?}
\author{Gilad Gour\thanks{E-mail:~gour@cc.huji.ac.il}$\;\;$ 
and$\;\;$L.~Sriramkumar\thanks{E-mail:~sriram@racah.phys.huji.ac.il}\\
Racah Institute of Physics, Hebrew University\\ 
Givat Ram, Jerusalem~91904, Israel.}

\date{}

\maketitle
\begin{abstract}
The Brownian motion of small particles interacting with a field at a 
finite temperature is a well-known and well-understood phenomenon. 
At zero temperature, even though the thermal fluctuations are absent,
quantum fields still possess vacuum fluctuations. 
It is then interesting to ask whether a small particle that is 
interacting with a quantum field will exhibit Brownian motion 
when the quantum field is assumed to be in the vacuum state. 
In this paper, we study the cases of a small charge and an imperfect
mirror interacting with a quantum scalar field in $(1+1)$~dimensions. 
Treating the quantum field as a classical stochastic variable, 
we write down a Langevin equation for the particles.
We show that the results we obtain from such an approach agree with 
the results obtained from the fluctuation-dissipation theorem.
Unlike the finite temperature case, there exists no special frame
of reference at zero temperature and hence it is essential that 
the particles do not break Lorentz invariance.
We find that that the scalar charge breaks Lorentz invariance,
whereas the imperfect mirror does not.
We conclude that small particles such as the imperfect mirror 
{\it will}\/ exhibit Brownian motion even in the quantum vacuum, 
but this effect can be so small that it may prove to be difficult 
to observe it experimentally. 
\end{abstract}
\newpage

\section{Introduction and motivation}

The random motion of a small particle that is immersed in a fluid
in thermal equilibrium is a phenomenon that has been known to us 
for a long time now.
No elaborate set up is required to observe this phenomenon. 
In fact, this phenomenon was first noticed by Brown, a botanist, 
early last century, when he observed, under a microscope, the 
motion of tiny pollen grains immersed in water at room temperature.
This random motion of small test particles in fluids, which has 
come to be known as Brownian motion, was originally understood on 
the basis of molecular or kinetic theory of fluids.
According to the kinetic theory, fluids consist of molecules which 
are in incessant and random motion because of intrinsic thermal 
fluctuations. 
Hence, if there is an external particle (which we shall hereafter 
refer to as the Brownian particle) present in a fluid, then the 
molecules of the fluid, apart from constantly colliding with each 
other, also collide with the Brownian particle, thereby imparting the 
particle with the observed random motion.
(For a good discussion on Brownian motion at finite temperature, 
see Pathria~\cite{pathria72}.)
Thus, Brownian motion reveals very clearly the statistical fluctuations
that occur in a system in thermal equilibrium. 
Historically, this phenomenon proved to be important in helping to gain 
the acceptance for the atomic theory of all matter and for the validity 
of the statistical description thereof.

Obviously, a Brownian particle cannot gain energy from the surrounding
medium indefinitely.
Therefore, there should exist a mechanism for the particle to 
dissipate its energy in some form so that it reaches equilibrium 
with the environment.
Early this century, it was Langevin who suggested that the force 
exerted on the Brownian particle by the surrounding medium can 
effectively be written as a sum of two parts:~(i)~an `averaged out' 
part which represents a frictional force experienced by the particle 
and (ii)~a `rapidly fluctuating' part~(see, for e.g., Reif~\cite{reif65}).
The `rapidly fluctuating' part is responsible for the random motion of
the Brownian particle and the presence of the frictional force implies 
the existence of processes whereby the energy associated with the 
Brownian particle is dissipated in course of time to the degrees of 
freedom corresponding to the surrounding medium.
Clearly, fluctuations and dissipation have to go hand in hand if the 
complete system has to stay in equilibrium.

A good example to illustrate the phenomenon we have described in
the last two paragraphs is the case of a test charge interacting 
with the electromagnetic field at a finite temperature.
According to quantum field theory, photons are the quanta of the 
electromagnetic field. 
The charge interacts with the photons present in the thermal bath 
and the recurrent collisions of the photons with the charge imparts 
the Brownian motion to the charge.
But, as we have pointed out in the last paragraph, a Brownian particle 
cannot keep accruing energy from the thermal fluctuations present in 
the surrounding environment. 
The charge, when in non-uniform motion, radiates photons and 
this radiation reacts back on the charge (see, for instance, 
Jackson~\cite{jackson62}, Chap.~17) with the result that it 
achieves the required equilibrium conditions.

Therefore, it is clear that dissipation of energy by a Brownian
particle is necessary to attain equilibrium in the presence of 
fluctuations.
In quantum statistical mechanics, the relation between 
fluctuation and dissipation is embodied in the 
fluctuation-dissipation theorem~\cite{caltwelt51}. 
(For a detailed account of the  fluctuation-dissipation
theorem, see Kubo~\cite{kubo66}.)
We had mentioned above that according to quantum field theory,
photons are the quanta of the electromagnetic field.
An important lesson we learn in quantum field theory is that 
the electromagnetic field has a non-zero energy, called the 
zero-point energy, even at zero temperature (see, for e.g., 
Milonni~\cite{milonni94}, Sec.~2.5.). 
In quantum statistical mechanics, we have come to understand that 
if the fluctuation-dissipation theorem has to be satisfied, we have 
to take the zero-point energy of the electromagnetic field into 
account (see, for e.g., Landau and Lifshitz~\cite{landau5}, Sec.~124).

The presence of the zero-point energy implies that fluctuations 
are present in the field even in the vacuum state (see, for instance, 
Milonni~\cite{milonni94}, Sec.~2.5.).
We had mentioned earlier that it is the presence of the fluctuations 
in the surrounding medium that is responsible for the random motion 
of the Brownian particles.
The question we are interested in addressing in this paper is as 
follows:~If fluctuations are present in a quantum field even in the 
vacuum state, then will a small particle that is interacting with the 
field exhibit Brownian motion in the quantum vacuum?
For the sake of simplicity, we shall study Brownian particles 
that are interacting with a quantized massless scalar field in 
$(1+1)$~dimensions.
We shall consider two kinds of Brownian particles:~(i)~a small 
scalar charge that is coupled to the field through a monopole 
interaction and (ii)~a mirror that is imperfect in the sense 
that it does not reflect modes higher than a certain frequency 
which we shall refer to as the plasma frequency.
 
This paper is organized as follows.
In Sec.~\ref{sec:laneqn}, treating the quantum field as a classical 
stochastic variable we write down a Langevin equation for the Brownian 
particles.
We shall consider the cases of a small charge and an imperfect 
mirror.
From the Langevin equation, we evaluate the mean-square velocities 
of these Brownian particles when they are in equilibrium with the
quantum field.
In Sec.~\ref{sec:fdt}, we compare the equilibrium values of the 
mean-square velocities we obtain from the Langevin equation 
with those obtained from the fluctuation-dissipation theorem.
In Sec.~\ref{sec:dffsn}, we evaluate the mean-square displacements 
of the small particles from the Langevin equation and examine whether 
the particles we consider will exhibit Brownian motion or not.
Finally, in Sec.~\ref{sec:dscsn}, we shall briefly summarize the  
main results of our analysis.
(Unless we mention otherwise, we shall work with units such that 
$\hbar=c=1$.)

\section{The Langevin equation}\label{sec:laneqn}

The systems we shall consider in this section are described by the 
action
\begin{equation}
{\cal S} = {\cal S}_{\rm par} + {\cal S}_{\rm fld}
+{\cal S}_{\rm int},\label{eqn:act}
\end{equation}
where ${\cal S}_{\rm par}$ represents the action corresponding to 
the Brownian particles, ${\cal S}_{\rm fld}$ denotes the action of
the field that the Brownian particles are interacting with and 
${\cal S}_{\rm int}$ is the action that describes the interaction 
between the Brownian particles and the field.
In the introductory section, we had mentioned that, for the sake 
of simplicity, we shall assume that the Brownian particles are 
interacting with a massless scalar field in $(1+1)$~dimensions.
In such a case, the action~${\cal S}_{\rm fld}$ is described by the 
integral
\begin{equation}
{\cal S}_{\rm fld} = \int dt\int dx\; 
\left\{\frac{1}{2}\, \partial^{\mu}\Phi\,\partial_{\mu}\Phi\right\},
\end{equation}
where $\Phi$~denotes the massless scalar field.
Also, we shall assume that the Brownian particles are moving 
non-relativistically. 
Then, the action ${\cal S}_{\rm par}$ describing a non-relativistic 
Brownian particle of mass~$m$ that is in motion (in $1$-dimension, 
say along the $x$-axis) on a trajectory $z(t)$ is given by
\begin{equation}
{\cal S}_{\rm par} 
= \int dt\; \frac{m}{2}\, {\dot z}^2, 
\end{equation}
where ${\dot z}\equiv(dz/dt)$. 

In the following two subsections, we shall study two kinds of Brownian 
particles interacting with the massless, quantum scalar field.
Treating the quantum field as a classical stochastic variable, we shall
write down a Langevin equation for the Brownian particles. 
In Subsec.~\ref{subsec:laneqnchrg}, we shall discuss the case of a 
small charge and, in Subsec.~\ref{subsec:laneqnmirr}, we shall 
consider the case of an imperfect mirror.
The explicit form of the interaction~${\cal S}_{\rm int}$ between the 
two kinds of Brownian particles and the massless scalar field will be 
given in the relevant subsection below.

\subsection{For a small scalar charge}\label{subsec:laneqnchrg}

In this subsection, we shall consider the case of a small charge
that is interacting with the massless scalar field through a monopole 
interaction. 
For such a case, the action ${\cal S}_{\rm int}$ is given by
\begin{equation}
{\cal S}_{\rm int} = \int dt\int dx\; \rho\,\Phi,\label{eqn:actintchrg}
\end{equation}
where $\rho$ is the charge density corresponding to the scalar charge.
As mentioned earlier, we shall assume that the Brownian particles are
moving non-relativistically.
The charge density $\rho$ corresponding to a non-relativistic charge 
moving along a trajectory $z(t)$ is given by
\begin{equation}
\rho(t,x)=q\; \delta^{(1)}\left[x-z(t)\right],\label{eqn:chrgdns}
\end{equation}
where $q$ is the strength of the scalar charge.
So, the complete system is now described by the action~(\ref{eqn:act}) 
with ${\cal S}_{\rm int}$ being given by Eqs.~(\ref{eqn:actintchrg}) 
and~(\ref{eqn:chrgdns}).
Varying the action~(\ref{eqn:act}) with respect to the scalar 
field~$\Phi$ and the trajectory~$z(t)$ of the charge, we find
that the equations of motion satisfied by the field and the 
charge are given by
\begin{equation}
\Box \Phi \equiv \left(\frac{\partial^2}{\partial t^2} 
-\frac{\partial^2}{\partial x^2}\right)\Phi
= q\; \delta^{(1)}\left[x - z(t)\right]\label{eqn:fldchrg}
\end{equation}
and
\begin{equation}
m\, {\ddot z} = q\; \left(\partial_{z(t)}
\Phi\left[t,{\bf z}(t)\right]\right),\label{eqn:chrg1}
\end{equation}
where 
\begin{equation}
{\ddot z}\equiv (d^2 z/dt^2)\qquad{\rm and}\qquad
\partial_{z(t)}\Phi\left[t,{\bf z}(t)\right]
\equiv \biggl({\partial \Phi\left[t,{\bf z}(t)\right]}
/{\partial z(t)}\biggl).
\end{equation}
In what follows, we shall first solve Eq.~(\ref{eqn:fldchrg}) for the 
scalar field and then substitute the resulting expression for~$\Phi$
in Eq.~(\ref{eqn:chrg1}) to obtain the final equation of motion for 
the charge.

The scalar field $\Phi$ satisfying Eq.~(\ref{eqn:fldchrg}) above 
can be decomposed as follows:
\begin{equation}
\Phi(t,x)=\Phi_{\rm free}(t,x)+\Phi_{\rm ret}(t,x),\label{eqn:decomp}
\end{equation}
where, as is obvious from the subscripts, $\Phi_{\rm free}$ and 
$\Phi_{\rm ret}$ denote the free and the the retarded components
of the scalar field, respectively.
The free component of the scalar field~$\Phi_{\rm free}$ satisfies 
the homogeneous wave equation and hence is independent of the charge 
density~$\rho$. 
Its most general solution can be written as a superposition of plane 
waves modes.
The retarded component of the field~$\Phi_{\rm ret}$ is a solution of 
the inhomogeneous wave equation. 
It is related to the charge density~$\rho$ by the following integral
(see, for e.g., Roman~\cite{roman69}, Sec.~3.1): 
\begin{equation}
\Phi_{\rm ret}(t, x) = \int\limits_{-\infty}^{\infty} dt'
\int\limits_{-\infty}^{\infty} dx'\; D_{\rm ret}(t,x;t',x')\;
\rho(t',x'),\label{eqn:phiret1}
\end{equation}
where $D_{\rm ret}$ is the retarded Green's function corresponding to 
the massless scalar field. 
In $(1+1)$~dimensions, the retarded Green's function $D_{\rm ret}$ 
can be easily evaluated to be (see, for instance, Birrell and 
Davies~\cite{bd82}, Sec.~2.7)
\begin{equation}
D_{\rm ret}(t,x;t',x')
=\theta(t-t')\, \int\limits_{-\infty}^{\infty} 
\frac{dk}{2\pi \omega}\; 
\sin\left[\omega(t-t')\right]\, 
e^{ik(x-x')},\label{eqn:retgfn}
\end{equation}
where $\omega=\vert k \vert$.
Substituting the expressions~(\ref{eqn:chrgdns}) 
and~(\ref{eqn:retgfn}) for the charge density~$\rho$ 
and the retarded Green's function~$D_{\rm ret}$ in 
Eq.~(\ref{eqn:phiret1}), we find that the retarded 
component of the scalar field reduces to the following 
integral:
\begin{equation}
\Phi_{\rm ret}(t,x) 
= q\, \int\limits_{-\infty}^{\infty} dt'\; \theta(t-t')
\int\limits_{-\infty}^{\infty} \frac{dk}{2\pi\omega}\, 
\sin\left[\omega(t-t')\right]\, 
e^{ik\left[x-z(t')\right]}.\label{eqn:phiret2}
\end{equation}

The decomposition of the scalar field~$\Phi$ into the free and 
the retarded components as in Eq.~(\ref{eqn:decomp}) leads to 
the following equation of motion for the charge:
\begin{equation}
m\, {\ddot z}= F_{\rm rr} + {\cal F},\label{eqn:chrg2}
\end{equation}
where
\begin{equation}
F_{\rm rr}= q\; \biggl({\partial}_{z(t)} 
\Phi_{\rm ret}\left[t,z(t)\right]\biggl)\qquad {\rm and}\qquad 
{\cal F}=q\; \biggl({\partial}_{z(t)} 
\Phi_{\rm free}\left[t,z(t)\right]\biggl).\label{eqn:frrcalf}
\end{equation}
The explicit form of~$F_{\rm rr}$ can now be obtained by 
substituting~$\Phi_{\rm ret}$ from Eq.~(\ref{eqn:phiret2}) 
in the above expression.
We find that~$F_{\rm rr}$ is described by the integral 
\begin{eqnarray}
F_{rr}&=&q\; \biggl({\partial_{z(t)}}\Phi_{\rm ret}
\left[t, z(t)\right]\biggl)\nonumber\\ 
&=& q^2 \int\limits_{-\infty}^{\infty}dt'\; \theta(t-t')
\int\limits_{-\infty}^{\infty} \frac{dk}{2\pi \omega}\; 
(ik)\;\sin\left[\omega(t-t')\right]\; 
e^{ik\left[z(t)-z(t')\right]}.
\end{eqnarray}
The integral over~$k$ can be expressed in terms of 
$\delta$-functions and as a result the intergal 
over~$t'$ can be easily evaluated.
We obtain that 
\begin{equation}
F_{\rm rr}= -\left(\frac{q^2}{2}\right) 
\left(\frac{\dot z}{1-{\dot z}^2}\right).\label{eqn:radreac}
\end{equation}
Recall that we had assumed that the charge is moving 
non-relativistically.
In the non-relativistic limit (i.e. when $\vert 
{\dot z}\vert \ll 1$), $F_{\rm rr}$ above reduces 
to $\left(-q^2 {\dot z}/{2}\right)$ with the 
result that the equation of motion for the charge 
is now given by
\begin{equation}
\frac{dv}{dt} +\left(\frac{q^2}{2m}\right) v
= \left(\frac{1}{m}\right){\cal F},\label{eqn:chrg3}
\end{equation}
where we have set $v={\dot z}$.
It is clear from this equation that the term~$F_{\rm rr}$ leads 
to dissipation.
$F_{rr}$, which arises from the retarded component of the field, 
is in fact the radiation reaction force on the scalar charge. 
Classically, it is possible to choose initial conditions such
that the free component of the scalar field is identically zero.
In such a case, ${\cal F}=0$ and the velocity of the charge 
decays to zero as the charge radiates when in motion.

Until now we have worked in the completely classical domain.
We have obtained an equation of motion for the charge assuming 
that the charge as well as the scalar field are classical 
quantities.
Our original motivation was to study the behavior of a Brownian 
particle that is interacting with a {\it quantum}\/ field.
If we now assume that~$\Phi$ is a quantum field, then 
the retarded component of the field, viz.~$\Phi_{\rm ret}$,
can still be regarded as a classical quantity, but the 
free component~$\Phi_{\rm free}$ should be treated as 
an operator (see, for e.g., Roman~\cite{roman69}, Sec.~3.1).
Therefore, on quantization of the scalar field, we would 
obtain an equation of motion for the charge that is similar 
in form to Eq.~(\ref{eqn:chrg3}), but the term~${\cal F}$ 
will now be an operator instead of a $c$-number.
In such a case, we will have an equation wherein the left 
hand side is a $c$-number whereas the right hand side is 
an operator and we need to devise an approach to make sense 
of such an equation.

One possible way out of this situation would be to replace 
the operator on right hand side by its expectation value.
As we are interested in studying motion in the quantum vacuum, 
the expectation value can be evaluated in the vacuum state of 
the quantum field.
In $(1+1)$~dimensions, the free component of the quantum scalar
field can be decomposed in terms of its normal modes as follows  
(cf.~Birrell and Davies~\cite{bd82}, Secs.~2.1 and~2.2):
\begin{equation}
{\hat \Phi}_{\rm free}(t,x)
=\int\limits_{-\infty}^{\infty} 
\frac{dk}{\sqrt{4\pi\omega}} 
\left({\hat a}_{k}\; 
e^{-i(\omega t -k x)} +{\hat a}_{k}^{\dag}\; 
e^{i(\omega t -k x)}\right),\label{eqn:phifop}
\end{equation}
where $\omega=\vert k\vert$ and ${\hat a}_{k}$ and 
${\hat a}_{k}^{\dag}$ are the annihilation and the 
creation operators corresponding to the mode $k$ of 
the quantum field.
Imposing the canonical commutation relations on the field 
and its conjugate momentum would lead to the standard 
commutation relation between the operators~${\hat a}_{k}$ 
and~${\hat a}_{k}^{\dag}$ (see, for e.g., Birrell and 
Davies~\cite{bd82}, Sec.~2.2).
The vacuum state~$\vert 0\rangle$ of the quantum scalar 
field is then defined as follows:
\begin{equation}
{\hat a}_{k} \vert 0\rangle = 0\qquad \forall k.
\end{equation}
Substituting Eq.~(\ref{eqn:phifop}) in the expression for 
${\cal F}$ in Eq.~(\ref{eqn:frrcalf}), we find that the 
corresponding operator is given by
\begin{eqnarray}
{\hat {\cal F}}\left[t,z(t)\right]
&\equiv& q\;\biggl(\partial_{z(t)}
{\hat \Phi}_{\rm free}\left[t,z(t)\right]\biggl)\nonumber\\
&=&q\; \int\limits_{-\infty}^{\infty} 
\frac{dk}{\sqrt{4\pi\omega}}\; (ik)\; 
\left({\hat a}_{k}\; 
e^{-i\left[\omega t -k z(t)\right]}  
- {\hat a}_{k}^{\dag}\; 
e^{i\left[\omega t -k z(t)\right]}\right).
\label{eqn:calfop}
\end{eqnarray}
From this expression, it is easy to see that its expectation 
value is zero in the vacuum state of the quantum field. 
Therefore, replacing the operator ${\hat {\cal F}}$ by its 
expectation value would simply lead us to the classical result 
and we will miss out the effects arising due to the quantum 
nature of the scalar field.

The main feature of a quantum field is that it always exhibits 
fluctuations.
This fluctuating nature of a quantum field induces fluctuations 
in the motion of the Brownian particles that are interacting 
with it.
Therefore, to study the effects of a quantum field on Brownian 
particles we need to formulate an approach wherein we are able 
to take into account the fluctuations that arise in the quantum 
field.
The approach we shall adopt here is to treat the force arising 
due to the quantum component of the scalar field as a classical 
stochastic force, say, $\beta(t)$, whose moments are related 
to the symmetrized $n$-point correlation functions of the operator 
${\hat {\cal F}}$.
In other words, we shall assume that the charge interacting with
the quantum scalar field is described by a Langevin equation of
the following form: 
\begin{equation}
\frac{dv}{dt}+ \gamma_{\rm c}\; v= \left(\frac{1}{m}\right)
\beta(t),\label{eqn:laneqnchrg}
\end{equation}
where \begin{equation}
\gamma_{\rm c}=\left(\frac{q^2}{2m}\right)\label{eqn:gammac}
\end{equation}
and $\beta(t)$ is a classical stochastic force whose first 
and second moments are given by 
\begin{eqnarray}
\langle \beta(t) \rangle &=& \biggl\langle {\hat {\cal F}}
\left[t,z(t)\right]\biggl\rangle_{\vert{\dot z}\vert\ll 1}
\label{eqn:firmom}\\
\langle \beta(t)\, \beta(t')\rangle 
&=& \left(\frac{1}{2}\right)\;
\biggl\langle{\hat {\cal F}}\left[t,z(t)\right]\;
{\hat {\cal F}}\left[t',z(t')\right]
+ {\hat {\cal F}}\left[t',z(t')\right]\;
{\hat {\cal F}}\left[t,z(t)\right]\biggl
\rangle_{\vert {\dot z}\vert \ll 1}.\qquad\label{eqn:secmom}
\end{eqnarray}
In these equations, the quantities on the left hand sides 
are to be considered as ensemble averages and the expectation 
values of the operators on the right hand sides are to be 
evaluated in the vacuum state of the quantum scalar field.
(In App.~\ref{app:props}, we show that the correlation 
functions we have defined here satisfy the required 
properties of a classical stochastic force.)
Moreover, since we have assumed that the charge is moving 
non-relativistically, it is necessary that we consider the 
$\vert{\dot z}\vert \ll 1$ limit of these expectation values. 

The presence of the stochastic force~$\beta(t)$ in the 
Langevin equation we have obtained above implies that 
quantities such as~$v(t)$ and~$z(t)$ that describe the 
motion of the charge exhibit fluctuations. 
Therefore, $v(t)$ and~$z(t)$ should be treated as 
stochastic variables.
We shall now solve Eq.~(\ref{eqn:laneqnchrg}) for~$v(t)$ and 
then go on to evaluate~$\langle v(t) \rangle$ and~$\langle 
v^2(t) \rangle$ by relating these quantities to  the first 
and the second moments of~$\beta(t)$.
We shall assume that the quantum scalar field is in the vacuum 
state.

Integrating the Langevin equation~(\ref{eqn:laneqnchrg}), we
obtain that
\begin{equation}
v(t)
=v(0)\; e^{-\gamma_{\rm c}t} + \left(\frac{1}{m}\right)\; 
e^{-\gamma_{\rm c}t}\; \int\limits_{0}^{t}dt\; 
e^{\gamma_{\rm c}t'}\; \beta(t'),\label{eqn:laneqnsoln}
\end{equation}
where we have set~$v(t=0)=v(0)$.
The expectation value of~$v(t)$ is then given by
\begin{eqnarray}
\langle v(t)\rangle
&=&\langle v(0)\rangle \; e^{-\gamma_{\rm c}t} 
+ \left(\frac{1}{m}\right)\; e^{-\gamma_{\rm c}t}\; 
\int\limits_{0}^{t}dt\; e^{\gamma_{\rm c}t'}\; 
\langle \beta(t') \rangle\nonumber\\
&=& v(0) \; e^{-\gamma_{\rm c}t}, 
\end{eqnarray}
where we have used the fact that the first moment of the
stochastic force~$\beta(t)$ is zero (cf.~App.~\ref{app:props}).
This is just the classical result we had discussed earlier.
The initial velocity~$v(0)$ of the charge decays to zero over
a time scale of the order of~$\gamma_{\rm c}^{-1}$.
It is then clear that the equilibrium 
value~$\langle v\rangle$ (i.e.~$\langle 
v (t) \rangle_{\gamma_{\rm c}t\gg1}$) 
is zero and the relaxation time of the 
system is~$\gamma_{\rm c}^{-1}$.
It is important to notice that there exists no special frame 
of reference at zero temperature.
Therefore, the fact that the equilibrium value of the velocity 
of the charge is zero implies that the scalar charge breaks 
Lorentz invariance~\cite{zurek86,unrzur89}.
(For a detailed discussion on this aspect, see 
App.~\ref{app:lorinv}.)

Let us now go on to evaluate the quantity~$\langle 
v^2(t) \rangle$.
Using Eq.~(\ref{eqn:laneqnsoln}), we obtain that
\begin{eqnarray}
\!\!\!\!\!\!\!\!\!\!\!\langle v^2(t) \rangle
= \langle v^2(0) \rangle \; e^{-2\gamma_{\rm c}t} 
&+& \left(\frac{2}{m}\right)\; e^{-\gamma_{\rm c}t}\;
\int\limits_{0}^{t}dt'\; e^{-\gamma_{\rm c}(t-t')}\; 
\langle v(0)\, \beta(t') \rangle\nonumber\\
&+& \left(\frac{1}{m^2}\right)\; e^{-2\gamma_{\rm c}t}\;
\int\limits_{0}^{t}dt'\; \int\limits_{0}^{t}dt''\; 
e^{\gamma_{\rm c}(t'+t'')}\; \langle \beta(t')\, 
\beta(t'') \rangle.\label{eqn:v2det}
\end{eqnarray}
If we now assume that~$t$ is large enough so 
that~$\left(\gamma_{\rm c}\,t\right)\gg 1$, then in such 
a limit the charge would be in equilibrium with 
the quantum scalar field.
In this limit, the first term in Eq.~(\ref{eqn:v2det})
clearly goes to zero and it can be shown that the 
second term reduces to zero as well\footnote{At a 
first glance, one would think that $\langle v(0)\,
\beta(t) \rangle$ is the same as $\left(v(0)\,\langle 
\beta(t) \rangle\right)$ and hence it should be 
identically zero. That is not the case. As we had 
pointed out earlier, the stochastic nature of $\beta(t)$ 
implies that $v(t)$ is a stochastic quantity as well. 
In general, there will exist non-zero correlations
between two stochastic variables. But, in the limit of
large~$t$, a correlation such as $\langle v(0)\, \beta(t) 
\rangle$ will decay exponentially (cf.~Landau and 
Lifshitz~\cite{landau5}, Secs.~118 and~119). 
Therefore, the integral in the second 
term is a finite quantity with the result that the coefficient 
$e^{-\gamma_{\rm c}t}$ kills this term completely in the 
large $t$ limit.}.
Therefore, we obtain that
\begin{equation}
\langle v^2 \rangle \equiv \langle v^2(t) 
\rangle_{\gamma_{\rm c} t \gg1}
=\left(\frac{\gamma_{\rm c}}{\pi m}\right) 
\int\limits_{0}^{\infty}
\frac{d\omega\; \omega}{\left(\gamma_{\rm c}^2
+\omega^2\right)},\label{eqn:v2chrg}
\end{equation}
where we have used the result~(\ref{eqn:beta2chrg}) 
for $\langle \beta(t')\, \beta(t'') \rangle$.
The expression for $\langle v^2\rangle$ we have obtained 
above diverges in the upper limit of the integral.
Such divergences are common in quantum field theory and
it is standard practice to regularize these divergences 
by introducing a finite upper limit to the integral, 
i.e.~by introducing an ultra-violet cut-off.
In the presence of the scalar charge, there exists a good 
reason to introduce such a cut-off. 
When the scalar charge is present, the quantum scalar 
field is obviously not a free field, but is interacting 
with the charge. 
If we now assume that the scalar charge has a finite 
size\footnote{Actually, the charge density as given in 
Eq.~(\ref{eqn:chrgdns}) corresponds to that of a 
point charge.
The charge density of a charge that has a finite size will be 
described by a distribution with a finite width rather than the
$\delta$-function.
The radiation reaction force on a charge of a finite size can 
be expressed as a power series in the size of the charge (see, 
for e.g., Jackson~\cite{jackson62}, Sec.~17.3). 
Therefore, by assuming that the radiation reaction 
force on the finite sized charge is still given by
Eq.~(\ref{eqn:radreac}), we are working under the 
approximation wherein we have retained only the 
leading order term.}, say, $\Lambda^{-1}$, then the 
modes of the quantum field with frequencies greater 
than~$\Lambda$ will not affect the charge.
The finite size of the charge will then provide us with 
a natural cut-off.
(It is for this reason that we have repeatedly mentioned 
that the scalar charge we are considering here is a small 
particle rather than a point particle.)
Carrying out the integral in Eq.~(\ref{eqn:v2chrg}) up 
to $\omega=\Lambda$, we obtain that
\begin{equation}
\langle v^2 \rangle 
=\left(\frac{\gamma_{\rm c}}{\pi m}\right) 
\int\limits_{0}^{\Lambda}\frac{d\omega\; 
\omega}{\left({\gamma_{\rm c}}^2+\omega^2\right)}
=\left(\frac{\gamma_{\rm c}}{2\pi m}\right)\,
\ln\left[1+\left(\frac{\Lambda^2}{\gamma_{\rm c}^2}
\right)\right].
\end{equation}
The only length scale available in the problem 
is~$\gamma_{\rm c}^{-1}$.
(In fact, $\gamma_{\rm c}^{-1}$ is the equivalent 
of the ``classical electron radius" for the case 
of the scalar charge we are considering here.)
Setting $\Lambda=\gamma_{\rm c}$, we finally 
obtain that
\begin{equation}
\langle v^2 \rangle 
=\left(\frac{\gamma_{\rm c}\, 
\ln 2}{2\pi m}\right).
\label{eqn:v2chrgf}
\end{equation} 
This result can be expressed as 
\begin{equation}
\frac{m}{2}\, \langle v^2 \rangle
=\left(\frac{\gamma_{\rm c}\, \ln 2}{4\pi}\right)\label{eqn:echrg}
\end{equation}
which is the average energy of the scalar charge when it is in 
equilibrium with the quantum scalar field. 
 
\subsection{For an imperfect mirror}\label{subsec:laneqnmirr}

We shall now study the case of a mirror that is interacting 
with the massless scalar field. 
In the last subsection, we had assumed that the charge was 
coupled to the scalar field through a monopole interaction
as described by the interaction term~(\ref{eqn:actintchrg}). 
In the case of the mirror, there exists no such explicit 
interaction term in the action (i.e. the quantity 
${\cal S}_{\rm int}$ is identically zero), but the mirror 
interacts with the field through a boundary condition.
We shall impose the boundary condition that the scalar field 
vanishes on the surface of the mirror.
In other words, we shall assume that
\begin{equation}
\Phi\left[t,z(t)\right]=0,\label{eqn:bc}
\end{equation}
where~$z(t)$ is the the trajectory of the mirror.
Since ${\cal S}_{\rm int}=0$, varying the action~(\ref{eqn:act})
with respect to~$\Phi$ leads to the following the equation of 
motion for the scalar field 
\begin{equation}
\Box \Phi \equiv \left(\frac{\partial^2}{\partial t^2} 
-\frac{\partial^2}{\partial x^2}\right)\Phi=0.
\end{equation}

The presence of the mirror implies that an incoming wave would
be reflected by the mirror into an outgoing wave.
If we assume that the scalar field~$\Phi$ is a classical field, 
then it is possible to choose initial conditions such that there 
are no incoming waves.
Even if there is an occasional incoming wave, scattering of such 
a wave by the mirror will just result in an impulsive change in 
the momentum of the mirror\footnote{Such a scattering would also 
result, due to momentum conservation, in a shift in the frequency 
of the scalar wave. This frequency shift will depend on the mass 
of the mirror and also on its velocity  at the point when it was 
hit by the incoming wave.}.
Moreover, this change in momentum (and hence the shift in the 
frequency of the reflected wave) will be very small if the mirror 
is assumed to be relatively heavy.
Apart from this effect, there would be no systematic effect of a 
classical scalar field on the motion of the mirror.
On the other hand, if we consider~$\Phi$ to be a quantum 
field, it is well-known that a mirror which is interacting 
with a quantum field can radiate even when the field is in 
the vacuum state~\cite{dewitt75,fulldav76}.
Such a radiation will then lead to a radiation reaction force on 
the mirror~\cite{fulldav76,fordvil82}.
This proves to be a feature that distinguishes mirrors from 
charges.
Unlike a charge which will radiate even when it is interacting 
with a classical field, a mirror will radiate only when it is 
interacting with a quantum field.  
 
In the absence of an explicit interaction term between the mirror
and the scalar field, varying the action~(\ref{eqn:act}) with 
respect to the trajectory~$z(t)$ of the mirror will just lead us 
to the equation of motion of a free particle.
In the last paragraph, we had pointed out that a mirror 
interacting with a quantum field will radiate when it is in motion.
This radiation will then lead to a non-zero radiation reaction 
force on the mirror, thereby affecting its motion.
Assuming~$\Phi$ to be a quantum field, we shall now obtain 
an equation of motion for the mirror by demanding that the total
energy of the system consisting of the mirror and the quantum 
field be conserved.

Conservation of energy implies that $\left(dH/dt\right)=0$, 
where $H$ is the Hamiltonian of the complete system.
We had noted above that the radiation reaction force on a 
moving mirror has a quantum origin.
If so, the radiation reaction force will exhibit fluctuations.
In order to take these fluctuations into account, we shall write 
the Hamiltonian of the complete system consisting of the mirror
and the quantum scalar field as follows:
\begin{equation}
H = H_{\rm mir} + {\hat H}_{\rm fld},\label{eqn:totham} 
\end{equation}
where $H_{\rm mir}$ denotes the Hamiltonian of the mirror
and ${\hat H}_{\rm fld}$ is the Hamiltonian operator of the 
scalar field.
The quantities $H_{\rm mir}$ and ${\hat H}_{\rm fld}$ are 
given by
\begin{equation}
H_{\rm mir}= \left(\frac{m}{2}\right)\; {\dot z}^2\qquad
{\rm and}\qquad {\hat H}_{\rm fld}
\equiv \int\limits_{-\infty}^{\infty} dx\;
{\hat T}_{00},\label{eqn:Hop}
\end{equation}
where, for the case of a massless scalar field in 
$(1+1)$~dimensions, the operator~${\hat T}_{00}$ 
is given by (cf.~Fulling and Davies~\cite{fulldav76})
\begin{equation}
{\hat T}_{00}=\left(\frac{1}{2}\right)
\left\{\left(\frac{\partial {\hat \Phi}}{\partial t}\right)^2
+ \left(\frac{\partial {\hat \Phi}}{\partial x}\right)^2\right\}.
\label{eqn:T00}
\end{equation} 
Demanding conservation of energy then leads to the following 
equation of motion for the mirror:
\begin{equation}
m\; {\dot z}\; {\ddot z} 
=- \left(d{\hat H}_{\rm fld}\left[t, z(t)\right]/dt\right).
\end{equation}
Adding $\left(d\langle{\hat H}_{\rm fld}\rangle/dt\right)$ to 
either side of this equation, we obtain that
\begin{equation}
m\; {\ddot z} -F_{rr}={\hat {\cal F}}.
\label{eqn:motmirr1}
\end{equation}
The quantities $F_{rr}$ and ${\hat {\cal F}}$ in the above 
equation are given by the expressions
\begin{equation}
F_{rr}= -\left[{\dot z(t)}\right]^{-1}\; 
\left(d\langle{\hat H}_{\rm fld}
\left[t, z(t)\right]\rangle/dt\right)\label{eqn:frrmirr}
\end{equation}
and
\begin{equation}
{\hat {\cal F}}[t, z(t)]=- \left[{\dot z(t)}\right]^{-1}\; 
\left(d{\hat {\cal H}}[t, z(t)]/dt\right),\label{eqn:calfmirr}
\end{equation}
where
\begin{equation}
{\hat {\cal H}}[t, z(t)]\equiv\left({\hat H}_{\rm fld}
\left[t, z(t)\right]-\langle{\hat H}_{\rm fld}
\left[t, z(t)\right]\rangle\right).
\label{eqn:calh}
\end{equation}
From the form of Eq.~(\ref{eqn:motmirr1}) it is easy to see 
that $F_{\rm rr}$ is the radiation reaction force on the mirror 
and ${\hat {\cal F}}$ represents the deviations of the radiation 
reaction force from its mean value.
(It should now be clear as to why we had considered 
the operator~${\hat H}_{\rm fld}$ rather than its 
expectation value in the total Hamiltonian of the
system as given by Eq.~(\ref{eqn:totham}). 
Had we considered the expectation value, we would have 
only obtained the term $F_{rr}$ and would have missed  
out the fluctuations arising due to the term~${\hat {\cal F}}$.)
As in the case of the charge, we shall treat the force 
on the mirror arising due to the term~${\hat {\cal F}}$  
as a classical stochastic force. 
We shall now calculate the radiation reaction force on the
mirror, viz.~$F_{\rm rr}$, from Eq.~(\ref{eqn:frrmirr}).
We shall assume that the quantum scalar field is in the vacuum 
state.

The boundary condition~(\ref{eqn:bc}) implies that the mirror 
separates the spacetime into two regions which are independent 
of each other.
An incoming wave in either of these regions is reflected by the 
mirror into an outgoing wave in the same region.
Let us now quantize the scalar field~$\Phi$ on either side of the 
mirror.
On the right hand side of the mirror, a positive frequency mode 
that satisfies the boundary condition~(\ref{eqn:bc}) is given by 
(cf.~Birrell and Davies~\cite{bd82}, Eq.~(4.43))
\begin{equation} 
\phi_{\omega}(t,x)=\frac{i}{\sqrt{4\pi\omega}}
\left[e^{-i\omega v}
-e^{-i\omega(2\tau_{u}-u)}\right],\label{eqn:moder}
\end{equation}
whereas such a mode on the left hand side of the mirror is 
given by
\begin{equation}
{\bar \phi}_{\omega}(t,x)=\frac{i}{\sqrt{4\pi\omega}}
\left[e^{-i\omega u}
-e^{-i\omega (2\tau_{v}-v)}\right],\label{eqn:model}
\end{equation}
where $\omega \ge 0$ and we have assumed that $z(t)=0$ for $t<0$.
The quantities $u$, $v$, $\tau_{u}$ and $\tau_{v}$ are defined 
by the relations
\begin{equation}  
u=(t-x),\quad v=(t+x),\quad \left[\tau_{u}-z(\tau_{u})\right]=u
\quad{\rm and}\quad \left[\tau_{v}+z(\tau_{v})\right]=v.
\end{equation}
(Note that the quantities $\tau_{u}$ and $\tau_{v}$ correspond to 
the time at which the incoming null waves $u$ and $v$ intersect 
the mirror, respectively.
Hence, $\tau_{u}$ depends only on~$u$ and $\tau_{v}$ only on $v$.)
The quantum scalar field can now be decomposed in terms of 
the modes~(\ref{eqn:moder})  and~(\ref{eqn:model}) as follows:
\begin{equation}
{\hat \Phi}^{\rm R}(t,x)=\int\limits_{0}^{\infty}d\omega\;
\biggl({\hat a}_{\omega}\; \phi_{\omega}(t,x)
+{\hat a}_{\omega}^{\dag}\; \phi_{\omega}^*(t,x)\biggl)
\label{eqn:phiopr}
\end{equation}
and 
\begin{equation}
{\hat \Phi}^{\rm L}(t,x)=\int\limits_{0}^{\infty}d\omega\;
\biggl({\hat b}_{\omega}\; {\bar \phi}_{\omega}(t,x)
+ {\hat b}_{\omega}^{\dag}\; {\bar \phi}_{\omega}^*(t,x)\biggl),
\label{eqn:phiopl}
\end{equation}
where the superscripts~${\rm R}$ and ${\rm L}$ denote the right 
and the left hand sides of the mirror, respectively.
The operators ${\hat a}_{\omega}$ and ${\hat a}_{\omega}^{\dag}$ 
(${\hat b}_{\omega}$ and ${\hat b}_{\omega}^{\dag}$) are the 
annihilation and the creation operators corresponding to the 
mode~$\omega$ on the right (left) hand side of the mirror. 
$\left({\hat a}_{\omega}, {\hat a}_{\omega}^{\dag}\right)$ and 
$\left({\hat b}_{\omega}, {\hat b}_{\omega}^{\dag}\right)$ form
two  independent sets of operators and operators within the same 
set satisfy the standard commutation relations.

Let us now evaluate the quantity $\left(d\langle{\hat H}_{\rm fld}
\rangle/dt\right)$ in the vacuum state of the quantum scalar field.
The vacuum states on either side of the mirror are 
defined as the states that are annihilated by the 
operators ${\hat a}_{\omega}$ and ${\hat b}_{\omega}$.
Substituting the scalar field in Eqs.~(\ref{eqn:phiopr}) 
and~(\ref{eqn:phiopl}) in the expression~(\ref{eqn:T00}) 
for ${\hat T}_{00}$, we find that its expectation value in 
the vacuum state on either side of the mirror  are given by 
\begin{equation}
\langle {\hat T}_{00}^{\rm R}\rangle
=\int\limits_{0}^{\infty}\frac{d\omega}{4\pi}\; \omega\; 
\left[1+(2{\dot \tau_{u}}-1)^2\right]
\qquad{\rm and} \qquad
\langle {\hat T}_{00}^{\rm L}\rangle
=\int\limits_{0}^{\infty}\frac{d\omega}{4\pi}\; \omega\; 
\left[1+(2{\dot \tau_{v}}-1)^2\right],\label{eqn:t00rt00l}
\end{equation}
where ${\dot \tau}_{u}\equiv\left(\partial 
\tau_{u}/\partial t\right)$ and ${\dot \tau}_{v}
\equiv\left(\partial \tau_{v}/\partial t\right)$ 
and these two quantities are related to the 
velocity of the mirror ${\dot z}(t)$ by the 
following relations:
\begin{equation}
{\dot \tau_{u}}=\left(\frac{1}{1-{\dot z(\tau_{u})}}\right)
\qquad{\rm and}\qquad
{\dot \tau_{v}}=\left(\frac{1}{1+{\dot z(\tau_{v})}}\right).
\end{equation} 
The integrals describing $\langle {\hat T}_{00}^{\rm R}\rangle$ and 
$\langle {\hat T}_{00}^{\rm L}\rangle$ in Eq.~(\ref{eqn:t00rt00l})
above exhibit the characteristic ultra-violet divergence of quantum
field theory.

The following few comments are in order at this stage of 
our discussion.
Earlier, in the case of the scalar charge, we had encountered 
an ultra-violet divergence when evaluating the equilibrium 
value of its mean-squared velocity, viz. the quantity
$\langle v^2\rangle$ (cf.~Eq.~(\ref{eqn:v2chrg})).
We can expect that such a divergence will arise for the case 
of the mirror as well.
However, as we have mentioned before, there exists an 
important difference between the case of the charge we 
had considered in the last subsection and the case of 
the mirror we are considering here.
The radiation reaction force on the scalar charge is classical 
in nature (cf.~Eq.~(\ref{eqn:frrcalf})), whereas the radiation 
reaction force on the mirror has a quantum origin  
(cf.~Eq.~(\ref{eqn:frrmirr})).
It is due to this reason that we encounter divergences 
even when we evaluate the radiation reaction force on 
the mirror.
In the case of the scalar charge, we had regularized the 
divergence in the mean-squared velocity by assuming that 
the charge has a finite size and then treating the finite 
size of the charge as an ultra-violet cut-off.
We can introduce such a cut-off for the mirror by assuming 
that it is imperfect in the sense that it does not reflect
modes higher than a certain frequency which we shall refer 
to as the plasma frequency.
Moreover, in order to be consistent, it is essential that we
evaluate not only the mean-squared velocity, but also the 
radiation reaction force on the mirror with the assumption 
that the mirror has a finite plasma frequency.

In what follows, we shall carry out our calculations assuming
that the mirror has a finite plasma frequency~$\omega_{\rm p}$.
In such a case, modes of the quantum field with frequencies 
$\omega>\omega_{\rm p}$ will be unaffected by the mirror and 
these modes would correspond to the standard Minkowski plane 
wave modes.
On subtracting the contribution to 
$\langle {\hat T}_{00}^{\rm R}\rangle$ and 
$\langle {\hat T}_{00}^{\rm L}\rangle$ due 
to the Minkowski vacuum for {\it all}\/ 
modes, we obtain that
\begin{equation}
\langle {\hat T}_{00}^{\rm R}\rangle 
=\int\limits_{0}^{\omega_{\rm p}}\frac{d\omega}{\pi}\;
\omega\;\left[{\dot \tau_{u}}\,({\dot \tau_{u}}-1)\right]
=\left(\frac{\omega_{\rm p}^2}{2\pi}\right) 
\left[{\dot \tau_{u}}\,({\dot \tau_{u}}-1)\right]
\end{equation}
and
\begin{equation}
\langle {\hat T}_{00}^{\rm L}\rangle
=\int\limits_{0}^{\omega_{\rm p}}\frac{d\omega}{\pi}\;
\omega\;\left[{\dot \tau_{v}}\,({\dot \tau_{v}}-1)\right]
=\left(\frac{\omega_{\rm p}^2}{2\pi}\right) 
\left[{\dot \tau_{v}}\,({\dot \tau_{v}}-1)\right].
\end{equation}
We had pointed out earlier that $\tau_{u}$ depends only on $u$
and $\tau_{v}$ only $v$. 
Therefore, the quantities $\langle{\hat T}_{00}^{\rm R}\rangle$ 
and $\langle {\hat T}_{00}^{\rm L}\rangle$ depend only on $u$ 
and $v$, respectively.
The expectation value of operator ${\hat H}_{\rm fld}$ 
in the vacuum state of the quantum scalar field is then 
given by
\begin{eqnarray}
\langle {\hat H}_{\rm fld}\rangle  
&=&\int\limits_{z(t)}^{\infty} dx\;  
\langle {\hat T}_{00}^{\rm R}(u)\rangle
+\int\limits_{-\infty}^{z(t)} dx\;
\langle {\hat T}_{00}^{\rm L}(v)\rangle \nonumber\\
&=&\int\limits_{-\infty}^{\left[t-z(t)\right]} du\; 
\langle {\hat T}_{00}^{\rm R}(u)\rangle 
+\int\limits_{-\infty}^{\left[t+z(t)\right]} dv\; 
\langle {\hat T}_{00}^{\rm L}(v)\rangle, 
\end{eqnarray}
where in the last equation we have changed the variables of 
integration from $x$ to $u$ and $v$.
From this expression it is easy to obtain that 
\begin{eqnarray} 
\left(\frac{d\langle {\hat H}_{\rm fld}\rangle}{dt}\right)
&=&\left[(1-{\dot z})\; \langle {\hat T}_{00}^{\rm R}(u)
\rangle_{\left[u=t-z(t)\right]}\right]
+\left[(1+{\dot z})\; \langle {\hat T}_{00}^{\rm L}(v)
\rangle_{\left[v=t+z(t)\right]}\right]\nonumber\\
&=&\left(\frac{\omega_{\rm p}^2}{\pi}\right) 
\left(\frac{{\dot z}^2}{1-{\dot z}^2}\right). 
\end{eqnarray}
In the non-relativistic limit, the radiation reaction 
force~$F_{rr}$ as given by Eq.~(\ref{eqn:frrmirr}) then
reduces to
\begin{equation}
F_{\rm rr}= -\left(\frac{\omega_{\rm p}^2\, {\dot z}}{\pi}\right).
\end{equation}
Substituting this expression in Eq.~(\ref{eqn:motmirr1}) and 
treating the force arising due to the term ${\hat {\cal F}}$ 
as a classical stochastic force $\beta(t)$, we find that the 
motion of the mirror is described by the Langevin 
equation~(\ref{eqn:laneqnchrg}) with $\gamma_{\rm m}$ instead 
of $\gamma_{\rm c}$, where $\gamma_{\rm m}$ is given by
\begin{equation}
\gamma_{\rm m}=\left(\frac{\omega_{\rm p}^2}
{\pi m}\right).\label{eqn:gammam}
\end{equation}
The first and the second moments of $\beta(t)$ are described 
by Eqs.~(\ref{eqn:firmom}) and~(\ref{eqn:secmom}) with the 
operator ${\hat {\cal F}}$ now given by Eq.~(\ref{eqn:calfmirr}).

The first and the second moments of $v(t)$ for the mirror can 
now be evaluated from the Langevin equation in the same fashion 
as in the case of the charge.
Just as in the case of the charge, the equilibrium 
velocity $\langle v\rangle$ (i.e. 
$\langle v(t)\rangle_{\gamma_{\rm m}t \ll 1}$) 
of the mirror is zero. 
But, unlike the case of the charge, such a behavior on the part 
of the mirror does not break Lorentz invariance.
The reason being that in obtaining this result we have assumed 
that the initial velocity of the mirror was zero.
(Recall that we had assumed that  $z(t)=0$ for $t<0$.)
Also, it can be shown that if the mirror was moving with 
a non-zero velocity initially then the equilibrium value 
$\langle v\rangle$ will be the same as the initial velocity.
(For details, see App.~\ref{app:lorinv}.)
The equilibrium value of $\langle v^2(t)\rangle$ for the 
imperfect mirror can now be obtained by substituting the 
second moment of the stochastic force as given 
by~(\ref{eqn:beta2mirrimp}) in Eq.~(\ref{eqn:v2det}). 
(The first term in Eq.~(\ref{eqn:v2det}) vanishes since 
the first moment of the stochastic force is zero for the 
mirror as well (see App.~\ref{app:props}) and the second 
term reduces to zero in the limit $\left(\gamma_{\rm m}\,
t\right)\gg1$ for the same reasons we had mentioned earlier.)
We find that 
\begin{equation} 
\langle v^2\rangle 
\equiv \langle v^2\rangle_{\gamma_{\rm m}t\gg1} 
=\left(\frac{1}{2\pi^2 m^2}\right)  
\int\limits_{0}^{\omega_{\rm p}}d\omega
\int\limits_{0}^{\omega_{\rm p}}d\omega' 
\left(\frac{\omega\omega'}{(\omega+\omega')^2
+\gamma_{\rm m}^2}\right).
\end{equation}
This integral can be evaluated by changing variables to 
$\Omega=\left[(\omega+\omega')/2\right]$ and $\Omega'
=\left[(\omega-\omega')/2\right]$ and carrying out the 
integrals over $\Omega$ and $\Omega'$.
We finally obtain that
\begin{eqnarray}
\!\!\!\!\!\!\!\!\!\!
{\langle v^2\rangle} 
&=&\left(\frac{\Gamma^2}{12}\right) 
\left\{3\; \ln\left(\frac{4+\Gamma^2}{1+\Gamma^2}\right)
+\Gamma^2\; \ln\left(\frac{\Gamma^4+ 4\, \Gamma^2}{1
+\Gamma^2}\right) -1\right\}\nonumber\\ 
& &\qquad\qquad\qquad\qquad\qquad-\;
\left(\frac{\Gamma}{3}\right)\left[{\rm arctan}(2\Gamma^{-1})
-{\rm arctan}(\Gamma^{-1})\right],
\label{eqn:lbutone} 
\end{eqnarray} 
where $\Gamma=\left(\gamma_{\rm m}/\omega_{\rm p}\right) 
=\left(\omega_{\rm p}/\pi m\right)$.  
In terms of $\hbar$ and $c$, the dimesionless quantity $\Gamma$ 
is given by 
\begin{equation}
\Gamma=\left(\frac{\hbar \omega_{\rm p}}{\pi m c^2}\right). 
\end{equation} 
A typical value for the plasma frequency $\omega_{\rm p}$ of
the mirror would be $10^{16}~sec^{-1}$ (cf.~Jackson~\cite{jackson62}, 
p.~321). 
For a mirror of mass $10^{-3}~kg$, $\Gamma\approx 10^{-31}$.
Clearly, $\Gamma \ll 1$. 
In this limit, we find that (\ref{eqn:lbutone}) reduces to
\begin{equation}
\langle v^2\rangle\simeq \left(\frac{\Gamma^2}{4}\right)\,
(\ln 4 -1)\simeq\left(\frac{\gamma_{\rm m}}{4\pi m}\right).
\label{eqn:v2mirrf}
\end{equation}
This expression can be rewritten as
\begin{equation}
\frac{m}{2}\, \langle v^2\rangle
=\left(\frac{\gamma_{\rm m}}{8\pi}\right)\label{eqn:emirr}
\end{equation}
which is the average energy of the mirror when it is in 
equilibrium with the quantum scalar field.

\section{Comparison with the fluctuation-dissipation 
theorem}\label{sec:fdt}

It is the surrounding medium that leads to the dissipative 
and the random forces on a Brownian particle.
Since these two forces have a common origin, they must be 
related in some fashion.
Fluctuation-dissipation theorem is a statement about a
general relationship between the response of a given system
to an external disturbance and the internal fluctuation of 
the system in the absence of this disturbance~\cite{caltwelt51}.
The fluctuation-dissipation theorem can be used to predict 
the fluctuations of physical quantities from the known
dissipative properties of the system when it is subject to 
an external interaction~\cite{kubo66}.
It is in this form that we shall use the theorem.

In this section, we shall first gather together the 
basic definitions and the essential results of the 
fluctuation-dissipation theorem.
Applying the fluctuation-dissipation theorem to the cases 
of the small charge and the imperfect mirror we evaluate 
the equilibrium values of the quantity $\langle v^2(t)\rangle$.
We shall then compare the results we obtain from the 
fluctuation-dissipation theorem with those we have 
obtained in the last section. 

The fluctuations of physical quantities can be related, 
in many cases, to quantities that describe the behavior 
of the body under certain external interactions.
{\it The fluctuating physical quantities can be either 
classical variables or quantum operators.}\/
The external interactions appear in the Hamiltonian of 
the body as a perturbation term of the following form  
(cf.~Landau and Lifshitz~\cite{landau5}, Sec.~123):
\begin{equation}
{\hat V}=-{\hat x}\;f(t),\label{eqn:linint}
\end{equation}
where~${\hat x}$ represents the physical quantity concerned 
and the perturbing generalized force~$f(t)$ is a given function 
of time.
The mean value of the quantity~ ${\hat x}(t)$ can be written as
\begin{equation}
\langle {\hat x(t)}\rangle
=\int\limits_{0}^{\infty}dt'\;\alpha(t')\;f(t-t'),
\label{eqn:resppert}
\end{equation}
where $\alpha(t')$ is a function of time which depends on 
the properties of the body.
It is clear from the above expression that the value 
of~$\langle {\hat x}\rangle$ at a time~$t$ depends only 
on the value of the perturbing force~$f$ at earlier times.
The quantity~$\langle {\hat x}\rangle$ is the response 
of the system to the perturbation.
Decomposing the quantities $\langle{\hat x}(t)\rangle$ 
and $f(t)$ in Eq.~(\ref{eqn:resppert}) in terms of their 
Fourier components $\langle {\hat x}_{\omega}\rangle$ and 
$f_{\omega}$, we find that they are related by
\begin{equation}
\langle {\hat x}_{\omega}\rangle = \alpha(\omega)\; f_{\omega},
\label{eqn:comp1}
\end{equation}
where the function~$\alpha(\omega)$ is described by the 
following integral:
\begin{equation}
\alpha(\omega)
=\int\limits_{0}^{\infty}dt\;\alpha(t)\;e^{i\omega t}.
\end{equation}
Once the function~$\alpha(\omega)$ is known, the behavior
of the body under the external perturbation is completely
determined.
The quantity~$\alpha(\omega)$ is called the generalized 
susceptibility and is, in general, a complex quantity.
We shall write~$\alpha(\omega)$ as
\begin{equation}
\alpha(\omega)=\alpha'(\omega)+i\; \alpha''(\omega)
\end{equation}
and the imaginary part~$\alpha''(\omega)$ charecterises 
the dissipative properties of the system (see Landau and 
Lifshitz~\cite{landau5}, p.~379).
Fluctuation-dissipation theorem essentially relates the 
equilibrium values of the fluctuations in the physical 
quantity~${\hat x}(t)$ and the dissipative properties of 
the system effectively represented by~$\alpha''(\omega)$.
The actual relationship between these quantities is given 
by the following expression (cf.~Landau and Lifshitz~\cite{landau5}, 
Eq.~(124.10)):
\begin{equation}
\langle {\hat x}^2 \rangle 
=\left(\frac{1}{\pi}\right)\int\limits_{0}^{\infty}d\omega\;
\alpha''(\omega)\; {\rm coth}(\omega/2T),
\end{equation}
where $T$~is the temperature of the surrounding medium.
At zero temperature, which is the case we are interested in,
the above relation reduces to
\begin{equation}
\langle {\hat x}^2 \rangle 
=\left(\frac{1}{\pi}\right)\int\limits_{0}^{\infty}d\omega\;
\alpha''(\omega).\label{eqn:x2fdt}
\end{equation}
{\it It should be emphasised here that the quantity~${\hat x}$
can correspond to either  a classical variable or a quantum 
operator.}

In the last section, we had obtained a Langevin to describe 
the motion of a small scalar charge and an imperfect mirror 
that are interacting with a quantum scalar field.
The Langevin equation we had obtained was of the 
following form (cf.~Eq.~(\ref{eqn:laneqnchrg})):
\begin{equation}
\frac{dv}{dt}+\gamma\; v
=\left(\frac{1}{m}\right)\beta(t),\label{eqn:laneqn}
\end{equation}
where $\gamma$ is the relaxation time of the system 
and $\beta(t)$ is a classical stochastic force.
We shall now identify the velocity~$v$ of the Brownian 
particles to be the physical quantity~${\hat x}$ that 
appears in our discussion on the fluctuation-dissipation 
theorem above.
Also, since~$\beta(t)$ is the force that induces fluctuations 
in the velocity~$v(t)$ of the Brownian particles, we shall 
identify the perturbing generalized force~$f(t)$ (as defined 
in Eq.~(\ref{eqn:linint})) with the stochastic force~$\beta(t)$. 
Comparing the dimensions of~$f(t)$ and~$\beta(t)$, we 
find that
\begin{equation}
\left[f(t)\right]=\left[\beta(t)\right]\cdot\left[sec\right],
\end{equation}
where the square brackets denote the dimensions of the 
quantities inside them.
The only time scale that is available in the problem
is~$\gamma^{-1}$.
Therefore, we shall assume that
\begin{equation}
\beta(t)\equiv\gamma\; f(t).
\end{equation}

We shall now calculate the generalized 
susceptibility~$\alpha(\omega)$ for the 
systems described by the Langevin 
equation~(\ref{eqn:laneqn}).
Expressing the quantities $v(t)$ and $\beta(t)$ in 
Eq.~(\ref{eqn:laneqn}) in terms of their Fourier 
components, we obtain that
\begin{equation}
m\; (\gamma-i\omega)\; v_{\omega}
=\beta_{\omega}=\gamma\; f_{\omega}.\label{eqn:comp2}
\end{equation}
Comparing Eqns.~(\ref{eqn:comp1}) and~(\ref{eqn:comp2}),
we find that~$\alpha(\omega)$ is given by
\begin{equation}
\alpha(\omega)=\left(\frac{\gamma}{m}\right)
\left(\frac{1}{\gamma-i\omega}\right)
=\left(\frac{\gamma}{m}\right)
\left(\frac{\gamma+i\omega}{\gamma^2+\omega^2}\right).
\end{equation}
It is now easy to identify that 
\begin{equation}
\alpha''(\omega)=\left(\frac{\gamma}{m}\right)
\left(\frac{\omega}{\gamma^2+\omega^2}\right).
\end{equation}
Substituting this expression for $\alpha''(\omega)$ in 
Eq.~(\ref{eqn:x2fdt}), we obtain that
\begin{equation}
\langle v^2 \rangle
=\frac{1}{\pi}\int\limits_{0}^{\infty}d\omega\; 
\alpha''(\omega)
=\left(\frac{\gamma}{m\pi}\right)
\int\limits_{0}^{\infty}
\frac{d\omega\; \omega}{(\gamma^2+\omega^2)}.
\end{equation}
This is exactly the intergal~(\ref{eqn:v2chrg}) we had 
encountered in the case of the scalar charge. 
This integral diverges in the upper limit and, as we have
discussed earlier, we can introduce a cut-off 
parameter~$\Lambda$ if we assume that the Brownian 
particles have a finite size $\Lambda^{-1}$.
Integrating up to $\Lambda$, we find that
\begin{equation}
\langle v^2\rangle
=\left(\frac{\gamma}{2\pi m}\right)\;
\ln\left[1+\left(\frac{\Lambda}{\gamma} \right)^2\right].
\end{equation}
Setting $\Lambda$ to be $\gamma_{\rm c}$ in the case of the
charge and $\omega_{\rm p}$ in the case of the mirror,  we 
find that
\begin{equation}
\langle  v^2\rangle_{\rm chr}
=\left(\frac{\gamma_{\rm c}\; 
\ln 2}{2\pi m}\right)\label{eqn:v2chrgfdt}
\end{equation}
and
\begin{equation}
\langle v^2\rangle_{\rm mir} 
=\left(\frac{\gamma_{\rm m}}{2\pi m}\right)\;
\ln\left[1+\left(\frac{\omega_{\rm p}}{\gamma_{\rm m}} 
\right)^2\right]
\simeq\left(\frac{\gamma_{\rm m}}{\pi m}\right)\; 
\ln(1/\Gamma),
\label{eqn:v2mirrfdt}
\end{equation}
where in the final expression we have used the fact that
$\Gamma\ll1$.

Let us now compare the results we have obtained from the 
fluctuation dissipation theorem with the results we had 
obtained in the last section.
Comparing Eq.~(\ref{eqn:v2chrgf}) with~(\ref{eqn:v2chrgfdt})
and Eq.~(\ref{eqn:v2mirrf}) with~(\ref{eqn:v2mirrfdt}), it 
is easy to see that the expressions match exactly in the case 
of the charge but agree only up to the leading order (in 
$\Gamma^2$) in the case of the mirror (for a detailed 
discussion on this issue, see Gour~\cite{gilad98}).
The difference that arises in the case of the mirror has a 
simple explanation.
The derivation of the fluctuation-dissipation 
theorem is crucially based on regarding the 
external interaction~(\ref{eqn:linint}) as a 
small perturbation which ensures that the 
response of the system is  linear (see Landau 
and Lifshitz~\cite{landau5}, p.~387).  
The interaction between the charge and the scalar field as
given by Eq.~(\ref{eqn:actintchrg}) is clearly linear.
But, the mirror interacts with the scalar field through a 
boundary condition and such an interaction is a complex one.  
Therefore, it does not come as a surprise that the result 
we have obtained in the last section for the case of the 
mirror agrees only up to the leading order with the result 
from the fluctuation-dissipation theorem.

\section{Will the small particles exhibit Brownian\\ 
motion?}\label{sec:dffsn} 

It is the time dependence of the mean-square displacement of
the small particles that reflects whether these particles will 
exhibit Brownian motion or not.
Earlier, we had obtained the Langevin equation~(\ref{eqn:laneqn})
to describe the motion of a small scalar charge and an imperfect 
mirror interacting with a quantum scalar field.
The mean-square displacement of the Brownian particles can 
be easily derived from the Langevin equation satisfied by
them (see, for e.g., Reif~\cite{reif65}).
We shall briefly discuss this derivation here in order to 
emphasize an assumption that will prove to be important for 
our discussion later on.

Multiplying the Langevin equation~(\ref{eqn:laneqn}) by~$z(t)$
and taking its mean value, we obtain that
\begin{equation} 
\biggl\langle z\,\frac{dv}{dt}\biggl \rangle 
+ \gamma\; \langle z v\rangle 
=\left(\frac{d\langle z\, v\rangle}{dt}\right) 
-\langle v^2\rangle  
+ \gamma\; \langle z\, v \rangle 
= \left(\frac{1}{m}\right)\langle z\, \beta(t)\rangle,
\end{equation}
where~$\langle v^2\rangle$ is the mean-square velocity of 
the Brownian particles when they are in equilibrium with
the quantum field.
The stochastic force is completely independent of the position 
of the Brownian particle.
Therefore,  we can set $\langle z(t)\, \beta(t)\rangle=0$.
On integrating the above differential equation twice and 
evaluating the constants of integration by assuming that 
$z(t=0)=0$, we find that the mean-square displacement of 
the Brownian particles is given by
\begin{equation} 
\langle  z^2(t)\rangle  
= 2 \gamma^{-1}  \langle v^2\rangle 
\left[t-\gamma^{-1}\left(1-e^{-\gamma t}\right)\right].
\end{equation}
The two limits of interest are 
$\left(\gamma\, t\right)\ll 1$ and 
$\left(\gamma\, t\right)\gg 1$.
When $\left(\gamma\, t\right)\ll 1$, the Brownian particle 
is yet to reach equilibrium with its environment. 
In this limit, the mean-square displacement of the 
particle is given by
\begin{equation}
\langle  z^2(t)\rangle_{\gamma t\ll 1} 
\approx\langle v^2\rangle\; t^2,
\end{equation}  
i.e. the particle behaves as a free particle moving 
with an average velocity $\sqrt{\langle v^2\rangle}$.
Whereas, when $\left(\gamma\, t\right)\gg 1$, the particle is 
in equilibrium with the field and, in this limit, we find that
\begin{equation}  
\langle z^2(t)\rangle_{\gamma t\gg 1} 
\approx 2\gamma^{-1} \langle v^2\rangle\; t. 
\end{equation}  
In other words, in an equilibrium situation the Brownian
particle diffuses through the surrounding medium.
Obviously, we need to know the typical value of the relaxation 
time of the system before we can say whether a Brownian particle 
will exhibit diffusion or not.

Until now, we have been treating the scalar charge and the mirror 
as classical Brownian particles. 
If we now assume that these Brownian particles are quantum objects,
then we can consider the Langevin equation we have obtained as a 
Heisenberg equation of motion of the following form:   
\begin{equation}  
\frac{d{\hat v}}{dt}+ \gamma_{\rm c}\; {\hat v}  
= \left(\frac{1}{m}\right){\hat {\cal F}},\label{eqn:laneqnop} 
\end{equation}  
where ${\hat v}$ is the velocity operator corresponding to the 
Brownian particles.
(Recall that, in the last section, we had emphasised that the 
quantity~${\hat x}$ can be treated as either a classical variable 
or a quantum operator.
Therefore, the results we have obtained and the conclusions we 
have drawn in the last two sections will hold good even if we 
treat the quantities describing the motion of the Brownian 
particles as operators rather than as classical variables.)
Earlier, when calculating the mean-square displacement of the 
classical Brownian particle we had assumed that $z(t=0)=0$.
If we now treat the Brownian particle as a quantum mechanical 
object we cannot set its position operator ${\hat z}(t)$ to be
identically zero at any instant of time because the position 
of the particle will always exhibit fluctuations.
Instead, we shall demand that  
\begin{equation}  
\langle{\hat z}(0)\, {\hat v}(0)
+{\hat v}(0)\, {\hat z}(0)\rangle=0, 
\end{equation}  
where the expectation value is evaluated in the vacuum state 
of the scalar field.  
(The Heisenberg equation of motion~(\ref{eqn:laneqnop}) relates
the operators describing the motion of the Brownian particle to 
those of the quantum scalar field.
The expectation values of the operators corresponding to the 
Brownian particles we consider here are evaluated in the vacuum 
state of the quantum scalar field.) 
Also, in the case of a quantum Brownian particle the initial 
uncertainty associated with the particle's position has to be 
take into account. 
Therefore, its mean-square displacement will be given by
\begin{equation} 
\langle  {\hat z}^2(t)\rangle  
= 2 \gamma^{-1}\langle {\hat v}^2\rangle 
\left[t-\gamma^{-1}\left(1-e^{-\gamma t}\right)\right] 
+\langle  {\hat z}^2(0)\rangle,\label{eqn:z2bp} 
\end{equation} 
where $\langle {\hat z}^2(0)\rangle$ is the uncertainty in the 
position of the particle at $t=0$.

It is now instructive to compare the above result for the quantum 
Brownian particles with the mean-square displacement of a free 
quantum particle. 
A free quantum particle satisfies the following Heisenberg 
equations of motion (cf.~Sakurai~\cite{sakurai94}):
\begin{equation}  
\frac{d{\hat z}}{dt}=\left(\frac{1}{m}\right)\; {\hat p}
\qquad{\rm and}\qquad \frac{d{\hat p}}{dt}=0.
\end{equation} 
These equations can be easily integrated to obtain the solution 
\begin{equation} 
{\hat z}(t)={\hat v}(0)t+{\hat z}(0),
\end{equation} 
where ${\hat v}(0)\equiv\left({\hat p}(0)/m\right)$.
Then, the mean-square displacement of the free particle
is given by
\begin{equation}
\langle {\hat z}^2(t)\rangle 
= \langle  {\hat v}^2(0)\rangle t^2
+ \langle  {\hat z}^2(0)\rangle,\label{eqn:z2fp}
\end{equation}
where, as in the case of the quantum Brownian particle, 
we have assumed that $\langle{\hat z}(0)\, {\hat v}(0) 
+{\hat v}(0)\, {\hat z}(0)\rangle=0$.
(The expectation values in the case of the free particle 
are assumed to be evaluated in a given state.)
In the limit of $\gamma \to 0$, we find that the mean-square 
displacement of a quantum Brownian particle as given by 
Eq.~(\ref{eqn:z2bp}) reduces to
\begin{equation} 
\langle  {\hat z}^2(t)\rangle
\stackrel{\gamma \to 0}{\longrightarrow}
\langle {\hat v}^2(0)\rangle t^2 +\langle  {\hat z}^2(0)\rangle 
\end{equation} 
which is the same as Eq.~(\ref{eqn:z2fp}).
This means that in the limit $\left(\gamma\, t\right)\ll1$ the 
intrinsic quantum nature of the Brownian particle dominates 
its motion and the particle behaves essentially as a free 
particle. 
By contrast, in the limit $\left(\gamma\, t\right)\gg 1$, the 
quantum nature of the field dominates the motion of the Brownian 
particle (as it is in equilibrium with the field) and the particle 
exhibits diffusion. 
(Note that, in the limit of large $t$, the initial uncertainty in 
the position of the Brownian particle, viz.~the quantity $\langle
{\hat z}^2(0)\rangle$ in Eq.~(\ref{eqn:z2bp}), can be neglected.)

Let us now examine whether the Brownian particles we have 
considered in this paper will exhibit diffusion or not.
In order to do so, we need to evaluate the numerical values 
of the relaxation time for the small charge and the imperfect 
mirror.
Introducing~$\hbar$ and~$c$ in the expressions 
(\ref{eqn:gammac}) and (\ref{eqn:gammam}) for 
$\gamma_{\rm c}$ and  $\gamma_{\rm m}$, we 
find that they are given by
\begin{equation}
\gamma_{\rm c} =\left(\frac{q^2}{2m c^2}\right)
\qquad {\rm and}\qquad
\gamma_{\rm m}
=\left(\frac{\hbar \omega_{\rm p}^2}{\pi m c^2}\right).
\end{equation}
In the case of the scalar charge, if we now assume that the 
magnitude of~$q$ is the same as that of  the electronic 
charge\footnote{It should be pointed here out that, unlike 
the electromagnetic charge which is a dimensionless quantity, 
the scalar charge $q$ we are considering here has dimensions 
of inverse time (in units such that $\hbar=c=1$). 
Therefore, the charge strength~$q$ and the electronic charge 
$e$ should actually be related by a parameter, say, $D$,  
which has the same dimensions as~$q$. 
We have assumed here that the magnitude of $D$ is order unity. 
The exact value of~$D$ can swing the value of $\gamma_{\rm c}$ 
either way.}\label{fn} 
and $m$ to be the mass of an electron, we find 
that $\gamma_{\rm c}^{-1}\approx 10^{25}~sec$.  
On the other hand, the age of the universe~$\tau$ is of the 
order of~$10^{17}~sec$.
Clearly,  $\gamma_{\rm c}^{-1}\gg \tau$. 
However, there exist two reasons that suggest that 
these estimates for the scalar charge should not be 
taken seriously.
Firstly, the ``classical electron radius'' of the scalar 
charge, viz.~$\left(c\, \gamma_{\rm c}^{-1}\right)$, 
corresponding to the above values turns out to be 
$\approx 10^{33}~m$! 
Secondly, the ``classical electron radius'' for a 
collection of these scalar charges turns out to be 
smaller than that of a single charge\footnote{These 
features of the scalar charge may be counter-intuitive,
but the origin of these features can be traced back to 
the fact that the charge strength~$q$ has non-zero dimensions.
It is first useful to note that the ``classical electron 
radius" of a unit charge interacting with the electromagnetic 
field in $(3+1)$~dimensions would be given by a quantity such 
as~$\gamma_{c}$ (see, for instance, Jackson~\cite{jackson62}, 
p.~790).
Such a quantity would be directly proportional to the square 
of the electronic charge and inversely proportional to the 
mass of the charge.
So, for a collection of~$n$ such electromagnetic charges, the 
``classical electron radius" would go as~$n$.
However, due to the fact that the charge strength~$q$ has 
non-zero dimensions, the ``classical electron radius" of 
the scalar charge we are considering here is given 
by~$\left( c\,\gamma_{\rm c}^{-1}\right)$ rather 
than~$\gamma_{\rm c}$.
Since $\left(c\,\gamma_{\rm c}^{-1}\right)$ is directly 
proportional to~$m$ and inversely proportional to~$q^2$, 
for a collection of~$n$ such scalar charges, 
$\left(c\, \gamma_{\rm c}^{-1}\right)$ goes as $(1/n)$.
This ``inverse dependence" (when compared with the 
electromagnetic case in $(3+1)$ dimensions) on the
mass and the charge strength is also quite likely to be 
the reason for the extraordinarily large value of the 
``classical electron radius" of a single scalar charge.}!!
Furthermore, as we have discussed earlier, the scalar 
charge breaks Lorentz invariance in the quantum vacuum.
For the case of the mirror, we had mentioned before that a 
typical value for its plasma frequency $\omega_{\rm p}$ would 
be~$10^{16}~sec^{-1}$. 
If we now assume that the mass of the mirror is very small,
say, $10^{-5}~kg$,\/ then  $\gamma_{\rm m}^{-1}\approx 
10^{13}~sec$,\/ which is smaller than $\tau$.
More importantly, unlike the case of the scalar charge, 
Lorentz invariance is preserved when the mirror is in 
motion in the quantum vacuum.
Therefore, we can conclude that small particles such as the 
imperfect mirror we have considered here {\it will}\/
exhibit Brownian motion in the quantum vacuum.

\section{Summary}\label{sec:dscsn} 

In this concluding section, we shall briefly summarize the main
results we have obtained in this paper.

First and foremost, we would like to emphasize again the crucial 
difference between motion at a finite temperature and motion
in the quantum vacuum.
At a finite temperature, the thermal bath provides a special 
reference frame, but no such frame exists at zero temperature.
Therefore, for a realistic system to exhibit Brownian motion 
in the quantum vacuum, it is absolutely essential that the 
system preserves Lorentz invariance.
Of the two systems we have considered in this paper, {\it since 
the scalar charge breaks Lorentz invariance whereas the mirror 
does not,}\/ we would like to emphasize here that the {\it the 
mirror is a more realistic example than that of the charge}.

Secondly, we would like to stress that our answer to the title
of this paper is in the affirmative.
{\it Small particles},\/ such as the imperfect mirror we have 
considered in this paper, {\it will, in principle, exhibit 
Brownian motion in the quantum vacuum}.\/
Since we find that the typical energy of the Brownian particles 
is of the order of~$\gamma$, we can, in fact, expect the following 
universal behavior of small particles in the quantum vacuum: 
\begin{equation} 
\langle {\hat z}^2(t)\rangle_{\gamma t\gg 1}\approx 
\left(\frac{\hbar}{m}\right)\; t\label{eqn:unvrsl}, 
\end{equation} 
where we have written $\hbar$ explicitly.
From this expression it is easy to see that it will take an 
object of mass $10^{-3}~kg$\/ about $10^{27}~sec$\/ to move 
through a distance of $10^{-2}~m$.  
Obviously, observing such a behavior experimentally will prove 
to be a difficult task. 
On the other hand, if we assume that a mirror can be constructed 
out of, say, $10^{3}$ atoms or so, then the mass of such a mirror 
would be about $10^{-24}~kg$. 
This mirror would diffuse through a distance of $10^{-2}~m$\/ 
within a rather short time scale of $10^{6}~sec$\/ (which is 
about a month long). 
Possibly, such an effect can be observed in the laboratory.

\section*{Acknowledgements} 
\noindent 
The authors would wish to thank Prof.~Jacob.~D.~Bekenstein 
for suggesting the problem and for his help during the 
course of this work. 
This work was supported in part by a grant from the Israel 
Science Foundation established by the Israel Academy of 
Sciences.  

\appendix

\section{Properties of the classical stochastic 
force}\label{app:props} 
 
In studying Brownian motion, it is usually assumed that the  
stochastic force that is responsible for the random motion of 
Brownian particles satisfies the following two properties (see 
Kubo~\cite{kubo66}, Sec.~3 or Saslaw~\cite{saslaw85}). 
(i)~A positive stochastic force is as probable as a negative one 
and therefore the first moment of a stochastic force should be 
identically zero.
(ii)~The correlation time of the perturbations is very short and
therefore the second moment of the stochastic force should be 
a sharply peaked function of the time interval between the two 
perturbations.
In this appendix, we shall show that the first and the second
moments of the stochastic force~$\beta(t)$  as we have 
defined in Eqs.~(\ref{eqn:firmom}) and~(\ref{eqn:secmom}) 
satisfy these two properties.
Note that the operator~${\hat {\cal F}}$  is given by 
Eq.~(\ref{eqn:calfop}) in the case of the charge and 
by Eq.~(\ref{eqn:calfmirr}) in the case of the mirror.
We shall discuss the case of the charge in App.~\ref{app:chrg}
and the case of the mirror in App.~\ref{app:mirr}.

\subsection{In the case of the scalar charge}\label{app:chrg}

In the discussion following Eq.~(\ref{eqn:calfop}), we 
had mentioned that the expectation value of the 
operator~${\hat {\cal F}}$ vanishes in the vacuum state 
of the quantum field.
This implies that $\beta(t)$ satisfies property~(i).

Let us now evaluate the second moment $\langle \beta(t)\, 
\beta(t')\rangle$ in the vacuum state of the scalar field.
We shall first evaluate the two point function $\langle
{\hat {\cal F}}\left[t, z(t)\right]\, {\hat {\cal F}}
\left[t', z(t')\right]\rangle$, take the $\vert {\dot z}
\vert \ll 1$ limit and then symmetrize the resulting 
quantity to finally obtain~$\langle \beta(t)\,\beta(t') 
\rangle$.
Using the expression~(\ref{eqn:calfop}) for~${\hat {\cal F}}$, 
it is easy to show that
\begin{eqnarray}   
\biggl\langle{\hat {\cal F}}\left[t, z(t)\right]\, 
{\hat {\cal F}}\left[t', z(t')\right]\biggl\rangle
&=& q^2 \int\limits_{0}^{\infty}\frac{dk}{4\pi}\;k\; 
\left(e^{-ik\left[t-t'-z(t)+z(t')\right]}
+ e^{-ik\left[t-t'+z(t)-z(t')\right]}\right)\nonumber\\
&=& -\left(\frac{q^2}{2\pi(t-t')^2}\right)
\Biggl\{\left(1+\frac{\left[z(t)-z(t')\right]^2}{(t-t')^2}
\right)\nonumber\\
& &\qquad\qquad\qquad\quad\times\;
\left(1-\frac{\left[z(t)-z(t')\right]^2}{(t-t')^2}
\right)^{-2}\Biggl\}.
\end{eqnarray}
The quantity $\left(\left[z(t)-z(t')\right]/(t-t')\right)$ 
appearing in the above expression can be considered to be the 
average velocity~${\dot z}$ of the Brownian particle between 
the two instants $t$ and $t'$. 
Then, in terms of~${\dot z}$
\begin{equation}  
\biggl\langle{\hat {\cal F}}\left[t, z(t)\right]\, 
{\hat {\cal F}}\left[t', z(t')\right]\biggl\rangle
= -\left(\frac{q^2}{2\pi(t-t')^2}\right)\;
\left(1+{\dot z}^2\right)\;\left(1-{\dot z}^2\right)^{-2}.
\end{equation}
If we now consider the~$\vert {\dot z}\vert \ll 1$ limit, 
this expression reduces to
\begin{equation}  
\biggl\langle{\hat {\cal F}}\left[t, z(t)\right]\, 
{\hat {\cal F}}\left[t', z(t')\right]\biggl\rangle_{\vert 
{\dot z}\vert \ll 1}
= -\left(\frac{q^2}{2\pi(t-t')^2}\right).
\end{equation}
This correlation function can be written in its integral form 
as follows:
\begin{equation}
\biggl\langle{\hat {\cal F}}\left[t, z(t)\right]\, 
{\hat {\cal F}}\left[t', z(t')\right]\biggl
\rangle_{\vert {\dot z}\vert \ll 1}
=\left(\frac{q^2}{2\pi}\right)\; \int\limits_{0}^{\infty}
d\omega\;\omega\; e^{-i\omega(t-t')}.
\end{equation}
(Compare this expression with the zero temperature limit 
of Eq.~$(1.2')$ in Caldeira and Legget~\cite{caldlegg83}.)
On symmetrizing this quantity with respect to $t$ and $t'$ and 
substituting the resulting expression in Eq.~(\ref{eqn:secmom}) 
we finally obtain that
\begin{eqnarray}
\langle \beta(t)\,\beta(t')\rangle
&=& \left(\frac{q^2}{4\pi}\right)\;
\int\limits_{0}^{\infty}d\omega\;
\omega\;\left(e^{-i\omega(t-t')}
+e^{i\omega(t-t')}\right)\\
&=& \left(\frac{q^2}{2\pi}\right)\; 
\int\limits_{0}^{\infty}d\omega\; 
\omega\; \cos\left[\omega(t-t')\right].
\label{eqn:beta2chrg}
\end{eqnarray}
This integral can be easily carried out with the result
\begin{equation}
\langle \beta(t)\,\beta(t')\rangle
=-\left(\frac{q^2}{2\pi(t-t')^2}\right).\label{eqn:uvdiv}
\end{equation}
As required by property~(ii), this is clearly a sharply 
peaked function of the time interval between the two 
perturbations\footnote{The reader may be puzzled by 
the overall minus sign that appears in the second moment  
of a classical stochastic force. The root cause of the minus 
sign are the integrals in Eq.~(\ref{eqn:beta2chrg}) which 
exhibit ultra-violet divergence. In field theory, ultra-violet 
divergences as in Eq.~(\ref{eqn:beta2chrg}) are handled by 
considering the quantity $(t-t')$ to be given by $(t-t'\pm 
i\epsilon)$, where $\epsilon\to 0^{+}$. Therefore, the 
second moment of the stochastic force should actually be 
written as
\begin{equation}
\langle \beta(t)\,\beta(t')\rangle
=-\left(\frac{q^2}{2\pi(t-t'\pm i\epsilon)^2}\right).
\end{equation}
This is a positive definite and infinite quantity in the 
limit $t\to t'$ provided we take this limit {\it before}\/ 
setting $\epsilon$ to be zero.}.

\subsection{In the case of the mirror}\label{app:mirr}

It is easy to see from Eq.~(\ref{eqn:calfmirr}) that the 
expectation value of ${\cal F}$ is identically zero. 
In other words, the first moment of the stochastic force 
$\beta(t)$ vanishes thereby satisfying property~(i) trivially.

In what follows, we shall first evaluate the second moment 
of the stochastic force~$\beta(t)$ for a perfect mirror.
Then, in the final expression, we shall restrict the upper
limit in the integrals over $\omega$ to $\omega_{\rm p}$ in 
order to obtain the results for the imperfect mirror. 
We had mentioned earlier that a perfect mirror divides the 
spacetime into two independent regions.
Therefore, the fluctuations on either side of the mirror are
completely independent.
We shall now evaluate the correlation function $\langle
{\hat {\cal F}}\left[t, z(t)\right]\, {\hat {\cal F}}
\left[t', z(t')\right]\rangle$ in the vacuum state on the 
right side of the mirror.
On substituting the expression for the scalar 
field~(\ref{eqn:phiopr}) in Eq.~(\ref{eqn:calh}) 
and regularizing the expectation values by subtracting 
the contribution due to the Minkowski vacuum, we find that
\begin{eqnarray}
& &\!\!\!\!\!\!\!\!\!\!\!\!\!\!\!\!
\biggl\langle{\hat {\cal H}}^{\rm R}[t, z(t)]\,
{\hat {\cal H}}^{\rm R}[t', z(t')]\biggl\rangle\nonumber\\
&=&\left(\frac{1}{16\pi^2}\right)\int\limits_{z(t)}^{\infty}dx
\int\limits_{z(t')}^{\infty}dx'\;
\Biggl\{\left[\int\limits_{0}^{\infty}d\omega\; \omega\;
\exp-i\omega(t-t'+x-x')\right]^2\nonumber\\
& &\qquad\qquad\qquad
+\;(2{\dot {\tau}_{u}}-1)^2\;
\left[\int\limits_{0}^{\infty}d\omega\; \omega\;
\exp-i\omega\left(t-t'-x-x'+2z({\tau}_{u})\right)\right]^2 
\nonumber\\
& &\qquad\qquad\qquad
+\;(2{\dot \tau'}_{u}-1)^2\;
\left[\int\limits_{0}^{\infty}d\omega\; \omega\;
\exp-i\omega\left(t-t'+x+x'-2z(\tau'_{u})\right)\right]^2 
\nonumber\\
& &\qquad\qquad\qquad
+\;(2{\dot \tau_{u}}-1)^2\;
(2{\dot \tau'}_{u}-1)^2\nonumber\\
& &\qquad\qquad\qquad
\times\;
\left[\int\limits_{0}^{\infty}d\omega\; \omega\;
\exp-i\omega\left(t-t'-x+x'+2\left[z({\tau}_{u})
-z(\tau'_{u})\right]\right)\right]^2\Biggl\}\nonumber\\
&=&\left(\frac{1}{16\pi^2}\right)\int\limits_{z(t)}^{\infty}dx
\int\limits_{z(t')}^{\infty}dx'\;
\Biggl\{\frac{1}{(t-t'+x-x')^4} 
+\frac{(2{\dot {\tau}_{u}}-1)^2}
{\left[t-t'-x-x'+2z({\tau}_{u})\right]^4}\nonumber\\
& &\qquad\qquad\qquad\qquad\qquad\quad\;\;\,
+\;\frac{(2{\dot \tau'}_{u}-1)^2}
{\left[t-t'+x+x'-2z(\tau'_{u})\right]^4}\nonumber\\
& &\qquad\qquad\qquad\qquad\qquad\quad\;\;\,
+\;\frac{(2{\dot \tau_{u}}-1)^2\;
(2{\dot \tau'}_{u}-1)^2}
{\left[t-t'-x+x'+2(z(\tau_{u})-z(\tau'_{u})\right]^4}\Biggl\}.
\end{eqnarray}
On differentiating this expression with respect to $t$ and $t'$ 
and dividing by the quantity $\left[{\dot z(t)}\,{\dot z(t')}
\right]$, we obtain that
\begin{eqnarray}
\biggl\langle{\hat {\cal F}}^{\rm R}\left[t, z(t)\right]\, 
{\hat {\cal F}}^{\rm R}\left[t', z(t')\right]\biggl\rangle
&=&\left(\frac{1}{16\pi^2}\right)
\left(\frac{\left[1+{\dot z}(t)\right]\left[1
+{\dot z}(t')\right]}
{\left[1-{\dot z}(t)\right]\left[1-{\dot z}(t')
\right]}\right)\nonumber\\
& &\qquad\quad
\times\; \left(\frac{1}{\left[t-t'+z(t)-z(t')\right]^4}
\right).\label{eqn:right}
\end{eqnarray}
Calculating in a similar fashion for the left hand side of the 
mirror, we obtain that
\begin{eqnarray}
\biggl\langle{\hat {\cal F}}^{\rm L}\left[t, z(t)\right]\, 
{\hat {\cal F}}^{\rm L}\left[t', z(t')\right]\biggl\rangle
&=&\left(\frac{1}{16\pi^2}\right)
\left(\frac{\left[1-{\dot z}(t)\right]
\left[1-{\dot z}(t')\right]}{\left[1+{\dot z}(t)\right]
\left[1+{\dot z}(t')\right]}\right)\nonumber\\
& &\qquad\quad
\times\;\left(\frac{1}{\left[t-t'-z(t)+z(t')\right]^4}\right).
\label{eqn:left}
\end{eqnarray}
Since the two sides of the mirror are independent of each other, 
the complete correlation function is a sum of the correlation 
functions on either side.
On adding the two correlation functions (\ref{eqn:right}) 
and (\ref{eqn:left}) and treating the quantity 
$\left(\left[z(t)-z(t')\right]/(t-t')\right)$ as the 
average velocity ${\dot z}$ of the mirror and finally 
taking the limit $\vert {\dot z}\vert\ll 1$, we obtain 
that
\begin{equation}  
\biggl\langle{\hat {\cal F}}\left[t, z(t)\right]\,  
{\hat {\cal F}}\left[t', z(t')\right]
\biggl\rangle_{\vert {\dot z}\vert\ll 1} 
=\left(\frac{1}{2\pi^2 (t-t')^4}\right). 
\end{equation} 
This correlation function can be represented by the following 
integral expression:
\begin{equation} 
\biggl\langle{\hat {\cal F}}\left[t, z(t)\right]\,  
{\hat {\cal F}}\left[t', z(t')\right]
\biggl\rangle_{\vert {\dot z}\vert\ll 1} 
=\left(\frac{1}{2\pi^2}\right) 
\int\limits_{0}^{\infty}d\omega\; \omega
\int\limits_{0}^{\infty}d\omega'\; \omega'
\; e^{-i(\omega+\omega')\,(t-t')}.
\end{equation} 
On symmetrizing this quantity with respect to $t$ and $t'$ and 
substituting the resulting expression in Eq.~(\ref{eqn:secmom}), 
we find that
\begin{equation} 
\langle\beta(t)\, \beta(t')\rangle 
=\left(\frac{1}{2\pi^2}\right) 
\int\limits_{0}^{\infty}d\omega\;\omega
\int\limits_{0}^{\infty}d\omega'\;\omega' 
\; \cos\left[(\omega+\omega')\,(t-t')\right]. 
\label{eqn:beta2mirr} 
\end{equation}
On integration, we get that
\begin{equation}
\langle\beta(t)\, \beta(t')\rangle 
=\left(\frac{1}{2\pi^2 (t-t')^4}\right),
\end{equation}
which is a very sharply peaked function of the time interval
between the two perturbations as required by property~(ii).
Now, restricting the upper limit of the integrals over $\omega$
and $\omega'$ to $\omega_{\rm p}$, we obtain that
\begin{equation}
\langle\beta(t)\, \beta(t')\rangle
=\left(\frac{1}{2\pi^2}\right)
\int\limits_{0}^{\omega_{\rm p}}d\omega\; \omega
\int\limits_{0}^{\omega_{\rm p}}d\omega'\;\omega'
\; \cos\left[(\omega+\omega')\, (t-t')\right]
\label{eqn:beta2mirrimp}
\end{equation}
which is the second moment of the stochastic force for the 
case of the imperfect mirror.

\section{Is Lorentz invariance preserved?}\label{app:lorinv} 

In this appendix, we shall first show as to how the mirror 
preserves Lorentz invariance whereas the scalar charge does 
not.
We shall then attempt to understand the origin of this 
difference in the motion of these two Brownian particles.

In Subsec.~\ref{subsec:laneqnchrg}, we had found that the 
initial velocity~$v(0)$ of the charge decays to zero over 
a period of time~$\gamma_{\rm c}^{-1}$.
This implies that, when in equilibrium, the mean velocity of 
the charge is zero.
Had we been working at a finite temperature, the thermal bath 
of quanta corresponding to the field would offer a special 
frame of reference and we can say that the mean velocity of 
the charge is zero with respect to this reference frame.  
But, at zero temperature, no such frame of reference exists and   
the fact that a charge moving with a uniform velocity radiates   
implies that Lorentz invariance is broken~\cite{zurek86,unrzur89}.  
 
However, as we had mentioned in Subsec.~\ref{subsec:laneqnmirr}, 
the fact that the mean velocity of the mirror when in equilibrium
is zero does not break Lorentz invariance because in obtaining 
this result we had assumed the initial condition that $z(t)=0$ 
for $t<0$.
By demanding that the total momentum of the system be conserved,
we shall now obtain the radiation reaction force on the mirror
when the initial velocity of the mirror is assumed to be~$v_{0}$.
The momentum of the complete system is given by
\begin{equation}
P=m\, {\dot z} + P_{\rm fld},\qquad{\rm where}\qquad
P_{\rm fld}=\int\limits_{-\infty}^{\infty}dx\; 
\langle {\hat T}^{10}\rangle\label{eqn:momentum}
\end{equation}
and ${\hat T}^{10}$ is given by (cf.~Fulling
and Davies~\cite{fulldav76})
\begin{equation}
{\hat T}^{10}=-\left(\frac{1}{2}\right)
\left\{\left(\frac{\partial {\hat \Phi}}{\partial t}\right)
\left(\frac{\partial {\hat \Phi}}{\partial x}\right)
+\left(\frac{\partial {\hat \Phi}}{\partial x}\right)
\left(\frac{\partial {\hat \Phi}}{\partial t}\right)\right\}.
\label{eqn:t10}
\end{equation}
(Since we are interested here only in the average velocity of 
the mirror, we shall not bother about the fluctuations that arise 
in the radiation reaction term.)
Momentum conservation then implies that the 
radiation reaction force on the mirror is 
given by~$F_{\rm rr}=-\left(dP_{\rm fld}/dt\right)$. 

The momentum-density of the field on the right hand side of the 
mirror, viz.~$\langle{\hat T}^{10}_{\rm R}\rangle$, can now be 
evaluated by substituting the scalar field~(\ref{eqn:phiopr}) 
in the expression for~${\hat T}^{10}$ above.
On assuming that the mirror has a finite plasma 
frequency~$\omega_{\rm p}$ and then subtracting 
the contribution due to the Minkowski vacuum for
all modes, we obtain that
\begin{equation} 
\langle{\hat T}^{10}_{\rm R}\rangle 
=\left(\frac{\omega_{\rm p}^2}{2\pi}\right) 
\left[{\dot \tau}_{u}({\dot \tau}_{u}-1)\right].
\end{equation}
The total momentum of the field to the right of the mirror is
then given by
\begin{eqnarray}
P_{\rm fld}^{\rm R} 
&=&\left(\frac{\omega_{\rm p}^2}{2\pi}\right) 
\int\limits_{z(t)}^{\infty}dx\, 
\left[{\dot \tau}_{u}({\dot \tau}_{u}-1)\right]\nonumber\\
&=&\left(\frac{\omega_{\rm p}^2}{2\pi}\right)
\left\{\int\limits_{z(t)}^{t}dx\, 
\left[{\dot \tau}_{u}({\dot \tau}_{u}-1)\right]
+ \int\limits_{t}^{\infty}dx\, 
\left[{\dot \tau}_{u}({\dot \tau}_{u}-1)\right]\right\}.
\end{eqnarray}
The reason for dividing this expression for~$P_{\rm fld}^{\rm R}$ 
into two integrals is due to the fact that $\tau_{u}>0$ for $u>0$.
On changing the variable of integration from~$x$ to~$u$ in
the first integral in the above expression, we obtain that
\begin{equation}
P_{\rm fld}^{\rm R} 
=\left(\frac{\omega_{\rm p}^2}{2\pi}\right)
\left\{\int\limits_{0}^{\left[t-z(t)\right]}du\, 
\left[{\dot \tau}_{u}({\dot \tau}_{u}-1)\right]
+\int\limits_{t}^{\infty}dx\, 
\left[{\dot \tau}_{u}({\dot \tau}_{u}-1)\right]\right\}.
\label{eqn:rfrnce}
\end{equation}
It is now easy to see from this expression that 
the first term corresponds to the case~$v_{0}=0$
(the case for which we had evaluated the radiation
reaction force earlier). 
Also, since ${\dot z}(\tau_{u})=v_{0}$ for $u<0$, 
we can set ${\dot \tau}_{u}=(1-v_{0})^{-1}$ in 
the second integral.
On differentiating $P_{\rm fld}^{\rm R}$ above 
with respect to~$t$, we obtain that
\begin{equation}
\left(\frac{dP_{\rm fld}^{\rm R}}{dt}\right)
=\left(\frac{\omega_{\rm p}^2}{2\pi}\right)
\left\{\left(\frac{{\dot z}}{(1-{\dot z})}\right)
-\left(\frac{v_{0}}{(1-v_{0})^2}\right)\right\}.
\end{equation} 
The quantity $\left(dP_{\rm fld}^{\rm L}/dt\right)$ on the 
left hand side of the mirror can be evaluated in a similar 
fashion.
We find that
\begin{equation}
\left(\frac{dP_{\rm fld}^{\rm L}}{dt}\right)
=\left(\frac{\omega_{\rm p}^2}{2\pi}\right)
\left\{\left(\frac{{\dot z}}{(1+{\dot z})}\right)
-\left(\frac{v_{0}}{(1+v_{0})^2}\right)\right\}.
\end{equation}
On adding these two quantities and neglecting terms of
order ${\dot z}^2$ and $v_{0}^2$, we finally obtain that
\begin{equation}
F_{\rm rr}=-\gamma_{\rm m}\;
\left({\dot z}-v_{0}\right).
\end{equation}
Such a radiation reaction force ensures that the 
average velocity of the mirror (viz. $\langle v(t)
\rangle_{\gamma_{\rm m} t\gg1}$) remains~$v_{0}$ 
at any later time with the result that Lorentz 
invariance is preserved.
In other words, the radiation reaction force does not 
affect the average velocity of the mirror, but only 
affects the fluctuations in the velocity. 

We shall now attempt to understand as to why the scalar
charge breaks Lorentz invariance, whereas the mirror does 
not.
Using dimensionality arguments, let us now construct the 
radiation reaction force on a charge when the strength of 
the charge is a {\it dimensionless}\/ quantity.
It is reasonable to assume that the radiation reaction force 
on the charge will not depend on the trajectory~$z(t)$, but 
only on its velocity~${\dot z}(t)$ and its derivatives, say, 
for instance, ${\ddot z}(t)$ and $\stackrel{\cdots}{z}(t)$.
It is then easy to show that the radiation reaction force 
on such a charge will be of the following form:
\begin{equation}
F_{\rm rr}= a({\dot z})\, {\ddot z}^2  
+ b({\dot z})\, \stackrel{\cdots}{z}, 
\end{equation}
where $a({\dot z})$ and $b({\dot z})$ are functions of the 
velocity~${\dot z}$.
(In the non-relativistic limit, we expect the functions 
$a({\dot z})$ and $b({\dot z})$ to reduce to constants 
of order unity\footnote{This is indeed what happens in 
the case of a charge interacting with the electromagnetic 
field in $(3+1)$~dimensions. In such a case, in the 
non-relativistic limit, $a({\dot z})=0$ and $b({\dot z})$ 
is a positive constant (see, for e.g.,  
Jackson~\cite{jackson62}, p.~784)}.)
Clearly, such a radiation reaction force will preserve Lorentz 
invariance.
However, as we have discussed earlier, the scalar charge 
we have considered possesses non-zero dimensions (see
footnote~\ref{fn}).
Evidently, the motion of the scalar charge breaks Lorentz 
invariance due to the fact that the charge strength~$q$ has 
non-zero dimensions. 

Let us now compare the cases of the charge and the mirror.
In the case of the scalar charge, the charge strength~$q$
(which is basically a coupling constant) appears 
{\it explicitly}\/ in the interaction term in the action 
describing the complete system.
Whereas, in the case of the mirror, there is no interaction
term in the action, but the mirror interacts with the field 
through a boundary condition.
Moreover, the plasma frequency~$\omega_{\rm p}$ (which acts 
as the coupling constant for the system) appears {\it only}\/ 
when we introduce it as an ultra-violet cut-off in order to 
regularize the divergent expressions and we do not expect 
the regularization procedure to change the physics involved.
These arguments suggest that Lorentz invariance may not be 
preserved whenever a coupling constant that possesses 
non-zero dimensions appears {\it explicitly}\/ in the 
interaction term in the action describing the complete 
system.


\newpage

\begin{thebibliography}{15}  
\bibitem{pathria72} 
R.~K.~Pathria, {\sl Statistical Mechanics}\/ (Pergamon Press,  
Oxford, 1972), Secs.~13.3 and~13.4. 
\bibitem{reif65} 
F.~Reif, {\sl Fundamentals of Statistical and Thermal Physics}\/  
(McGraw-Hill, New York, 1965), pp.~560--567.
\bibitem{jackson62}
J.~D.~Jackson, {\sl Classical Electrodynamics},\/ Second Edition 
(Wiley, New York, 1962).
\bibitem{caltwelt51}
H.~B.~Callen and T.~A.~Welton, {\sl Phys.\ Rev.}\ {\bf 83}, 34 (1951).
\bibitem{kubo66}
R.~Kubo, {\sl Rep.\ Prog.\ Phys.}\ {\bf 29}, 255 (1966).
\bibitem{milonni94}
P.~W.~Milonni, {\sl The Quantum Vacuum}\/ 
(Academic Press, Boston, 1994).
\bibitem{landau5}
L.~D.~Landau and E.~M.~Lifshitz, {\sl Statistical Physics},\/ 
Part~I (Course of Theoretical Physics, Vol.~5), Third Edition 
(Pergamon Press, Oxford, 1980).
\bibitem{roman69}
P.~Roman, {\sl Quantum Field Theory}\/ (Wiley, New York, 1969).
\bibitem{bd82}
N.~D.~Birrell and P.~C.~W.~Davies, {\sl Quantum Fields in Curved 
Space}\/ (Cambridge University Press, Cambridge, England, 1982).
\bibitem{zurek86}
W.~H.~Zurek, {\sl Ann.\ N.\ Y.\ Acad.\ Sci.}\ {\bf 480}, 89 (1986).
\bibitem{unrzur89}
W.~G.~Unruh and W.~H.~Zurek, {\sl Phys.\ Rev.\ D}\ {\bf 40}, 
1071 (1989).
\bibitem{dewitt75}
B.~DeWitt, {\sl Phys.\ Reps.}\ {\bf 19C}, 297 (1975).
\bibitem{fulldav76}
S.~A.~Fulling and P.~C.~W.~Davies, {\sl Proc.\ Roy.\ Soc.\ Lond.\ A} 
{\bf 348}, 393 (1976).
\bibitem{fordvil82}
L.~H.~Ford and A.~Vilenkin, {\sl Phys.\ Rev.\ D}\ {\bf 25}, 
2569 (1982).
\bibitem{gilad98}
Gilad Gour, {\sl Motion in the Quantum Vacuum},\/ M.~Sc. Thesis,   
Hebrew University, Jerusalem, Israel (1998).
\bibitem{sakurai94}
J.~J.~Sakurai, {\sl Modern Quantum Mechanics}\/ 
(Addison-Wesley, Reading, Massachusetts, 1994), 
pp.~84--87.
\bibitem{milonni81}
P.~W.~Milonni, {\sl Phys.\ Letts.\ A}\ {\bf 82}, 225 (1981).
\bibitem{saslaw85}
W.~C.~Saslaw, {\sl Gravitational Physics of Stellar and Galactic 
Systems}\/ (Cambridge University Press, Cambridge, England, 1985), 
Chap.~3. 
\bibitem{caldlegg83}
A.~O.~Caldeira and A.~J.~Legget, {\sl Physica}\ {\bf 121~A}, 587
(1983). 
\end{thebibliography}
\end{document}